\documentclass[journal,twoside]{IEEEtran}
\usepackage[normalem]{ulem}
\usepackage{cancel}
\usepackage{tabularx,booktabs}
\usepackage{lipsum}
\newcolumntype{C}{>{\centering\arraybackslash}X} 

\usepackage{cite}

\usepackage[font=footnotesize]{caption}
\usepackage[font=footnotesize]{subcaption}
\usepackage{wrapfig}
\ifCLASSINFOpdf
   \usepackage[pdftex]{graphicx}
   \graphicspath{{../pdf/}{../jpeg/}}
   \DeclareGraphicsExtensions{.pdf,.jpeg,.png}
\else
   \usepackage[dvips]{graphicx}
   \graphicspath{{../eps/}}
   \DeclareGraphicsExtensions{.eps}
\fi
\usepackage[cmex10]{amsmath}
\usepackage{textcomp,gensymb}
\usepackage{relsize}

\usepackage{scalerel} 
\usepackage{tikz} 
\usetikzlibrary{svg.path}

\definecolor{orcidlogocol}{HTML}{A6CE39}
\tikzset{
	orcidlogo/.pic={
		\fill[orcidlogocol] svg{M256,128c0,70.7-57.3,128-128,128C57.3,256,0,198.7,0,128C0,57.3,57.3,0,128,0C198.7,0,256,57.3,256,128z};
		\fill[white] svg{M86.3,186.2H70.9V79.1h15.4v48.4V186.2z}
		svg{M108.9,79.1h41.6c39.6,0,57,28.3,57,53.6c0,27.5-21.5,53.6-56.8,53.6h-41.8V79.1z M124.3,172.4h24.5c34.9,0,42.9-26.5,42.9-39.7c0-21.5-13.7-39.7-43.7-39.7h-23.7V172.4z}
		svg{M88.7,56.8c0,5.5-4.5,10.1-10.1,10.1c-5.6,0-10.1-4.6-10.1-10.1c0-5.6,4.5-10.1,10.1-10.1C84.2,46.7,88.7,51.3,88.7,56.8z};
	}
}

\newcommand{\orcidicon}[1]{\href{https://orcid.org/#1}{\mbox{\scalerel*{
				\begin{tikzpicture}[yscale=-1,transform shape]
				\pic{orcidlogo};
				\end{tikzpicture}
			}{|}}}}

\usepackage [hidelinks] {hyperref} 

\usepackage{algorithmic}
\usepackage[ruled,vlined]{algorithm2e} 
\usepackage{amsmath,amssymb,amsfonts}

\begin{document} 
\title{Differential LEO Navigation under Asynchronous Satellite Clocks: Architecture and Performance Bounds}

\author{Qamar~Bader\textsuperscript{\orcidicon{0000-0002-4667-1710}}\,,~\IEEEmembership{Member,~IEEE,}
        Sharief~Saleh\textsuperscript{\orcidicon{0000-0003-1365-417X}}\,,~\IEEEmembership{Member,~IEEE,} 
        Henk Wymeersch\textsuperscript{\orcidicon{0000-0002-1298-6159}}\,,~\IEEEmembership{Fellow,~IEEE,}
        Gonzalo Seco-Granados\textsuperscript{\orcidicon{0000-0003-2494-6872}}\,,~\IEEEmembership{Fellow,~IEEE,}
        and   Aboelmagd~Noureldin\textsuperscript{\orcidicon{0000-0001-6614-7783}}\,,~\IEEEmembership{ Fellow,~IEEE}

\thanks{This work was supported by grants from the Natural Sciences and Engineering Research Council of Canada (NSERC) under grant numbers: ALLRP-560898-20 and RGPIN-2020-03900, and in part by the Spanish R+D project PID2023-152820OB-I00 and the Catalan AGAUR-ICREA Academia Program. (\textit{Corresponding author: Qamar~Bader.})}

\thanks{Qamar Bader and Aboelmagd Noureldin are with the Department of Electrical and Computer Engineering, Queen's University, Kingston, ON K7L 3N6, Canada, and also with the Navigation and Instrumentation (NavINST) Lab, Department of Electrical and Computer Engineering, Royal Military College of Canada, Kingston, ON  K7K 7B4, Canada (e-mail: qamar.bader@queensu.ca; aboelmagd.noureldin@rmc.ca).}
\thanks{Sharief Saleh and Henk Wymeersch are with the Department of Electrical Engineering, Chalmers University of Technology, SE 41296 Gothenburg, Sweden (e-mail: sharief@chalmers.se; henkw@chalmers.se).}
\thanks{Gonzalo Seco-Granados is with the Department of Telecommunications and Systems Engineering, Universitat Autonoma de Barcelona, 08193 Barcelona, Spain (e-mail: gonzalo.seco@uab.cat).}

\thanks{Digital Object Identifier 10.1109/TITS.2024.3480525}}

\markboth{IEEE Transactions on XXX XXX XXX}%
{Bader \MakeLowercase{\textit{et al.}}: Differential LEO Navigation under Asynchronous
Satellite Clocks: Architecture and Performance
Bounds}



\IEEEpubid{\begin{minipage}[c]{\textwidth}\ \\[12pt]
		 0018-9545 \copyright 2024 IEEE. Personal use of this material is permitted. Permission from IEEE must be obtained for all other uses, in any current or future media, including reprinting/republishing this material for advertising or promotional purposes, creating new collective works, for resale or redistribution to servers or lists, or reuse of any copyrighted component of this work in other works
	\end{minipage}}


\maketitle
\bstctlcite{IEEEexample:BSTcontrol} 

\begin{abstract}
Low Earth Orbit (LEO) communication satellites have emerged as a promising source of signals of opportunity for resilient positioning, navigation, and timing (PNT) in GNSS-challenged environments. Unlike GNSS constellations, however, commercial LEO systems are not globally synchronized, and independent satellite clock biases and drifts severely degrade kinematic estimation and induce statistical inconsistency when not explicitly managed. This paper proposes a base-station-aided differential navigation architecture that explicitly mitigates independent LEO satellite clock biases and drifts without inflating the state vector of the mobile rover. A fixed base station continuously tracks time-varying per-satellite clock states and transmits them as measurement-domain corrections—along with their rigorously propagated uncertainties—to a compact, 8-state rover Extended Kalman Filter (EKF). To support this architecture, we introduce a robust visibility management scheme to seamlessly handle the frequent entry and exit of LEO satellites. Furthermore, we derive the Recursive Bayesian Cramér-Rao Bound (RBCRB) directly linked to the channel-domain signal model to establish fundamental theoretical performance limits. The proposed methodology is validated using real-world urban vehicular data combined with high-fidelity LEO constellation simulations. Results demonstrate that the differential framework completely eliminates the statistical inconsistency endemic to single-receiver baselines. Crucially, the rover's posterior estimation uncertainty perfectly tracks the theoretical RBCRB limits across all kinematic and clock states, ensuring highly reliable PNT even during periods of severe geometric degradation.
\end{abstract}


\section{Introduction}
\IEEEPARstart{L}{ow} Earth orbit (LEO) communication satellites are increasingly viewed as an attractive source of signals of opportunity for positioning, navigation, and timing (PNT), particularly in environments where conventional GNSS becomes unreliable due to blockage and non-line-of-sight conditions \cite{11303905,Stock2024SurveyLEOSoOP}. In LEO-based PNT, the fundamental observables utilized to constrain receiver position, velocity, and clock states are the propagation delay (time-of-flight) and the Doppler shift \cite{khalife2021navigation,10388052}. Until recently, Doppler measurements have been significantly more accessible from uncooperative commercial constellations, leading to a large body of Doppler-only navigation frameworks \cite{fang2026robust,TRANSITonSteroids,shi2023revisiting,Time-DiverseDoppler}. However, recent breakthroughs in reverse-engineering the downlink signal structures of broadband mega-constellations such as Starlink \cite{Humphreys,UnveilingStarlink}, and OneWeb \cite{komodromos2026signal} have enabled the precise extraction of delay observables. This capability has catalyzed a shift toward joint delay-Doppler positioning architectures, which offer vastly improved geometric and clock observability compared to Doppler-only systems. However, unlike GNSS, LEO communication constellations are not primarily designed to provide a globally disciplined time scale for external PNT users. For instance, empirical analyses of the Starlink frame clock demonstrate that it exhibits discontinuous adjustments and lacks the stability of GNSS clocks \cite{qin2025analysis}. As a result, residual timing and frequency offsets generally differ across satellites and cannot be captured by a single common transmit-clock term. Consequently, unmodeled satellite timing errors can severely degrade the navigation filter, either through weak observability (if satellite clocks are appended to the receiver state) \cite{khalife2021navigation} or through filter inconsistency (if they are ignored) \cite{shi2022leo}. Accordingly, modeling each visible satellite with its own clock bias and drift becomes essential.
\IEEEpubidadjcol  
Early signals-of-opportunity (SoOP) demonstrations showed that \emph{Doppler-only} observations from LEO communication constellations can support receiver positioning/velocity estimation, but these formulations are typically single-epoch and treat satellite timing terms as nuisance effects---often via simplifying assumptions that do not explicitly manage per-satellite clock errors \cite{Tan2019IridiumSoOP,Guo2023InstantDopplerLEO}.  

Differential SoOP navigation has also been investigated, utilizing a known base station to form differenced observables that suppress common-mode errors and enhance system robustness \cite{Neinavaie2022DiffDopplerStarlink,SaroufimLEODiff,saroufim2024analysis,jiangDD,10012370}. Nevertheless, these methodologies strictly assume short baselines so that baseline-dependent atmospheric effects (ionospheric and tropospheric delays) perfectly cancel out, effectively isolating the clock dynamics in the differenced observables \cite{10320306}. Nevertheless, these methodologies exhibit fundamental limitations. First, they strictly assume short baselines so that baseline-dependent atmospheric effects perfectly cancel out. Second, they typically rely on direct measurement differencing to algebraically eliminate the remaining satellite clock errors at each isolated epoch, a mathematical cancellation that completely bypasses the rapid temporal dynamics and severe stochastic drift inherent to low-cost, uncooperative LEO oscillators. The work in \cite{yang2020real} demonstrated the real-time estimation of LEO satellite clocks using a wide-area network of ground tracking stations, ultimately intending to broadcast these states to augment traditional GNSS precise point positioning (PPP). While they explicitly estimate individual LEO clocks, their analysis primarily focuses on accelerating the convergence time of GNSS-assisted stationary receivers, rather than enabling standalone, LEO-only navigation for dynamic receivers. Furthermore, while standard GNSS augmentation protocols (such as RTCM SSR) broadcast deterministic corrections paired with scalar variance indicators (e.g., user range accuracy (URA)), these 1D bounds are insufficient to capture the highly coupled delay-Doppler drift of uncooperative LEO cheap oscillators. This necessitates a shift toward fully stochastic, covariance-aware correction mechanisms.

To properly validate any proposed LEO navigation architecture, it must be benchmarked against absolute information-theoretic limits. While an ample body of literature has utilized the Cramér-Rao lower bound (CRLB) to quantify how geometry and measurement quality constrain achievable accuracy \cite{8957111,li2025instant, jiangDD}, these existing frameworks exhibit critical gaps. First, they are often posed at the abstract measurement or state level, rather than being explicitly tied to a channel-domain signal model that produces link-dependent delay and Doppler information \cite{8957111}. Second, they are predominantly restricted to static, snapshot scenarios \cite{li2025instant,jiangDD}. While recursive Bayesian bounds are well-established in general distributed tracking theory \cite{tichavsky1998posterior}, to the best of the authors' knowledge, the LEO navigation literature currently lacks any theoretical framework that quantifies the impact of a tracked prior, specifically the stochastic injection of a tracked clock-state covariance from a ground reference station, on the rover's fundamental error limits.

In this paper, we address the problem of estimating a moving-target states (e.g., position, velocity, and receiver clock bias/drift) under asynchronous LEO satellite clocks in a differential positioning framework using delay and Doppler measurements. Addressing this problem is useful for achieving statistically consistent fusion and gaining deeper insights into fundamental tracking performance in dynamic satellite visibility conditions. The contributions of this paper can be summarized as follows:
\begin{enumerate}
  \item \textbf{Base-station-aided architecture for real-time LEO clock tracking:} We propose a differential framework tailored specifically for standalone SoOP-based LEO navigation. By offloading the tracking of asynchronous LEO clock states to a fixed base station, our architecture enables real-time positioning for dynamic rovers while maintaining a robust and computationally efficient 8-state kinematic extended Kalman filter (EKF).
  
  \item \textbf{Covariance-aware measurement-domain correction protocol:} We introduce a mathematically rigorous measurement-domain correction protocol to integrate the base station's clock estimates with the rover's kinematic filter. By formally propagating the per-satellite $2 \times 2$ posterior covariance matrix of the base station's estimates into the rover's measurement update, the framework explicitly mitigates filter overconfidence.

  \item \textbf{Formulation of theoretical performance benchmarks for dynamic and asynchronous LEO navigation:} To properly evaluate the proposed architecture, we derive the system-specific analytical Fisher information matrix (FIM) necessary to establish the recursive Bayesian Cram\'er-Rao bound (RBCRB). Within this formulation, we explicitly quantify the impact of the tracked prior, specifically the injection of the base station's stochastic clock covariance, on the rover's information matrix. Through high-fidelity kinematic simulations driven by real-world urban vehicular trajectories, we demonstrate that our correction protocol uniquely maintains strict statistical consistency, converging tightly with these derived theoretical limits.
\end{enumerate}

The remainder of the paper is structured as follows: Section~\ref{sec:System Model} outlines the considered scenario, signal model, and geometric relations. Section~\ref{Methodology} details the proposed differential navigation architecture, including the base station and rover estimators. Section~\ref{sec:crlb} derives the instantaneous and recursive Bayesian Cram\'er--Rao bounds. Section~\ref{sec:road test setup} furnishes details about the experimental and road test setup. Section~\ref{sec:results} presents the results and discussions. Finally, Section~\ref{sec:conclusion} concludes the paper.

\section{System Model} \label{sec:System Model}
\subsection{Scenario}
\begin{figure}
    \centering
    \includegraphics[width=\linewidth]{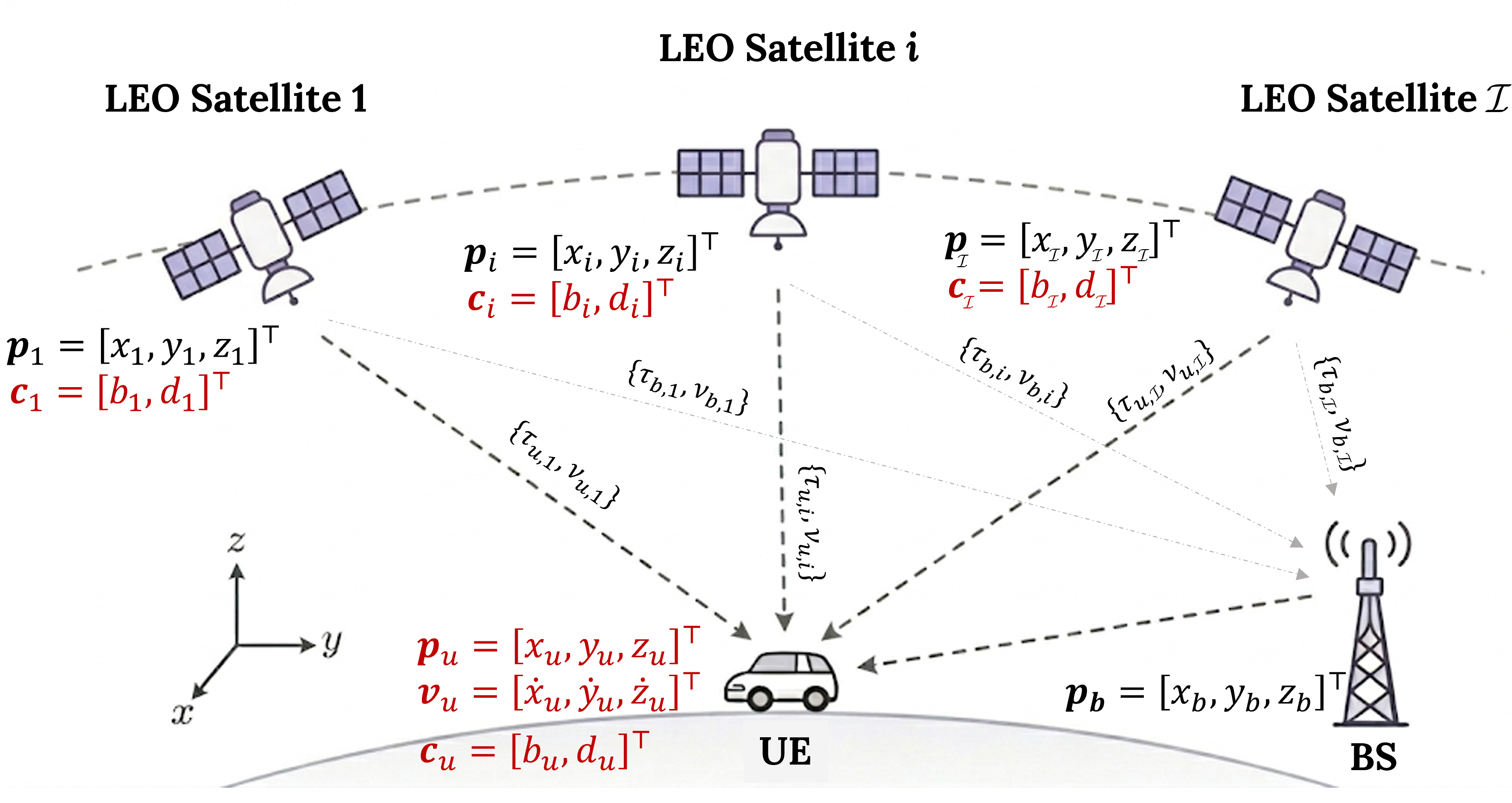}
    \caption{Illustration of the proposed differential LEO navigation scenario. A stationary BS and a dynamic UE simultaneously acquire joint delay ($\tau$) and Doppler ($\nu$) observables from $\mathcal{I}$ uncooperative LEO satellites.}
    \label{fig:scenario}
\end{figure}

We consider a differential positioning scenario involving a moving rover (mobile station) and a nearby fixed base station (BS), both simultaneously observing downlink signals from a set of visible LEO satellites indexed by $i\in\mathcal{I}_k$ at discrete epochs $k \in K$, as depicted in Fig.~\ref{fig:scenario}. The sequential epochs occur at physical times $t_k = k \Delta t$, where $\Delta t$ is the sampling interval of the estimation framework. The rover state is described by its position and velocity vectors, $\mathbf{p}_{u,k}\in\mathbb{R}^3$ and $\mathbf{v}_{u,k}\in\mathbb{R}^3$, respectively, in an Earth-centered Earth-fixed (ECEF) frame, together with a receiver clock vector $\mathbf{c}_{u,k} = [b_{u,k}\;\; d_{u,k}]^\top$ incorporating the clock bias $b_{u,k}$ and clock drift $d_{u,k}$. The BS has a known position $\mathbf{p}_b$, and a clock that is tightly disciplined to the universal/global GNSS time reference. Consequently, the BS clock bias and drift are effectively zero ($b_{b,k} = 0, d_{b,k} = 0$), and any residual hardware jitter is absorbed into the measurement noise. Each satellite $i$ has a known position and velocity, $\mathbf{p}_{i,k}$ and $\mathbf{v}_{i,k}$, respectively, and an unknown satellite clock vector $\mathbf{c}_{i,k}= [b_{i,k}\;\; d_{i,k}]^\top$. All clock terms are defined with respect to a universal/global clock reference.

All transceivers (rover user equipment (UE), BS, and satellites) are assumed to employ single antennas; consequently, no antenna orientation/attitude states are modeled. Throughout, we assume line-of-sight (LoS) propagation for the retained links and a time-varying satellite visibility set $\mathcal{I}_k$; consequently, satellites may enter or exit the estimation process across epochs. Unless stated otherwise, atmospheric and multipath effects are neglected, and all clocks evolve according to stochastic models specified in the sequel.
\subsection{Signal and Channel Model}
We assume that each LEO satellite employs downlink orthogonal frequency-division multiplexing (OFDM) and transmits known pilot symbols over an observation window of $L$ consecutive OFDM symbols indexed by $l\in\{0,\dots,L-1\}$, and $N$ subcarriers indexed by $n\in\{0,\dots,N-1\}$. Let $\Delta f$ denote the subcarrier spacing and $T=1/\Delta f$ the useful symbol duration. With a cyclic prefix duration $T_{\text{CP}}$, the OFDM symbol duration is $T_{\text{sym}}=T+T_{\text{CP}}$.

We model the satellites as being frequency-multiplexed, where each satellite $i\in\mathcal{I}_k$ occupies a dedicated, non-overlapping subset of subcarriers $\mathcal{N}_{i,k}\subset\{0,\dots,N-1\}$ (e.g., frequency-division multiple access (FDMA) allocation). The known transmitted pilot on subcarrier $n$ and OFDM symbol $l$ from satellite $i$ is denoted by $x^{n,l}_i\in\mathbb{C}$, where $x^{n,l}_i=0$ for $n \notin\mathcal{N}_{i,k}$.

For receiver $r\in\{u,b\}$, the frequency-domain complex baseband received signal on subcarrier $n$ and OFDM symbol $l$ at epoch $k$ is modeled as
\begin{equation}\label{eq:rx_signal}
y^{n,l}_{r,k}=\sum_{i\in\mathcal{I}_k} H^{n,l}_{r,i,k}\,x^{n,l}_i+\omega^{n,l}_{r,k},
\end{equation}
where $H^{n,l}_{r,i,k}$ is the frequency-domain channel response for the link between satellite $i$ and receiver $r$, and $\omega^{n,l}_{r,k}\sim\mathcal{CN}(0,\sigma_r^2)$ is circularly-symmetric complex AWGN where $\sigma_r^2=N_0 N_f$ is receiver noise variance, $N_0$ is the receiver thermal noise power spectral density (PSD), and $N_f$ is the receiver noise figure. We adopt a wideband single-path (dominant-path) model for each satellite link over the $L$-symbol window, for a given epoch $k$, expressed as (see Eq.~(21) in \cite{sallouha2024ground})
\begin{equation}\label{eq:channel_model}
H^{n,l}_{r,i,k}=\alpha_{r,i,k}\,
e^{-j2\pi n\Delta f\big(\tau_{r,i,k}-\nu_{r,i,k}\,l\,T_{\text{sym}}\big)}\,
e^{j2\pi f_c\,\nu_{r,i,k}\,l\,T_{\text{sym}}},
\end{equation}
where, $\alpha_{r,i,k}\in\mathbb{C}$ denotes the complex channel gain, $f_c$ is the carrier frequency, $\tau_{r,i,k}$ is the effective propagation delay over the observation window, and $\nu_{r,i,k}$ is the Doppler factor. The model in \eqref{eq:channel_model} captures the delay-induced linear phase progression across subcarriers and the Doppler-induced phase evolution across both OFDM symbols and subcarriers.
\subsection{Geometric Relations}
The channel parameters in \eqref{eq:channel_model} are governed by the satellite--receiver geometry and the involved clock states. The effective delay is modeled as
\begin{equation}\label{eq:tau_relation}
\tau_{r,i,k}=\frac{\big\|\mathbf{p}_{i,k}-\mathbf{p}_{r,k}\big\|}{c}+b_{r,k}-b_{i,k},
\end{equation}
where $\big\|\mathbf{p}_{i,k}-\mathbf{p}_{r,k}\big\|$ denotes the geometric range between the receiver position $\mathbf{p}_{r,k}$ and the satellite position $\mathbf{p}_{i,k}$, and $c$ is the speed of light.

\noindent The Doppler factor is modeled as
\begin{equation}\label{eq:nu_relation}
\nu_{r,i,k}= -\frac{\mathbf{u}_{r,i,k}^\top\big(\mathbf{v}_{i,k}-\mathbf{v}_{r,k}\big)}{c}+d_{r,k}-d_{i,k},
\end{equation}
where,
\begin{equation}\label{eq:los_vector}
\mathbf{u}_{r,i,k}= \frac{\mathbf{p}_{i,k}-\mathbf{p}_{r,k}}{\big\|\mathbf{p}_{i,k}-\mathbf{p}_{r,k}\big\|}
\end{equation}
is the corresponding LoS unit vector, pointing from the receiver $r$ to the transmitter $i$, and $\mathbf{v}_{r,k}$ and $\mathbf{v}_{i,k}$ denote the receiver and satellite velocities, respectively.

\subsection{Clock Dynamics Model}
\label{subsec:sys_clock_model}

To characterize the asynchronous timekeeping across the LEO constellation and the receiver, each local oscillator is modeled as a two-state stochastic dynamic system comprising a time (phase) bias state $b_k$ and a frequency (drift) state $d_k$.  For a discrete sampling interval $\Delta t$, the stochastic evolution of the clock errors is governed by a first-order Gauss--Markov (i.e., AR(1)) process for the frequency drift, which is subsequently integrated into the time bias
\begin{align}
d_{k} &= \phi \, d_{k-1} + \sigma_{y} \, w_{k}, \label{eq:gm-drift} \\
b_{k} &= b_{k-1} + d_{k-1} \, \Delta t + \sigma_{b} \, v_{k}, \label{eq:bias-update}
\end{align}
where $\phi = \exp(-\Delta t / \tau)$ represents the discrete-time correlation coefficient corresponding to the oscillator's correlation time $\tau$. The terms $w_{k}, v_{k} \sim \mathcal{N}(0,1)$ represent independent standard Gaussian driving sequences. Consequently, $\sigma_{y}$ and $\sigma_{b}$ dictate the standard deviations of the discrete-time frequency and time noise increments, respectively.

\subsection{Problem Formulation}
Given the received pilot observations \eqref{eq:rx_signal} over a time-varying visibility set $\mathcal{I}_k$, both the rover and the BS extract per-satellite delay and Doppler measurements $\{\tau_{r,i,k},\nu_{r,i,k}\}_{i\in\mathcal{I}_k}$ governed by \eqref{eq:tau_relation}--\eqref{eq:nu_relation}. The objective is to estimate the rover navigation and clock state $\{\mathbf{p}_{u,k},\mathbf{v}_{u,k},\mathbf{c}_{u,k}\}$ over time while accounting for unknown and time-varying satellite clock states $\{\mathbf{c}_{i,k}\}_{i\in\mathcal{I}_k}$ and the frequent satellite entry/exit typical of LEO dynamics. 

\section{Proposed Differential Navigation Architecture} \label{Methodology}
\subsection{Methodology Overview}\label{subsec:method_overview}
\begin{figure}
    \centering
    \includegraphics[width=1\linewidth]{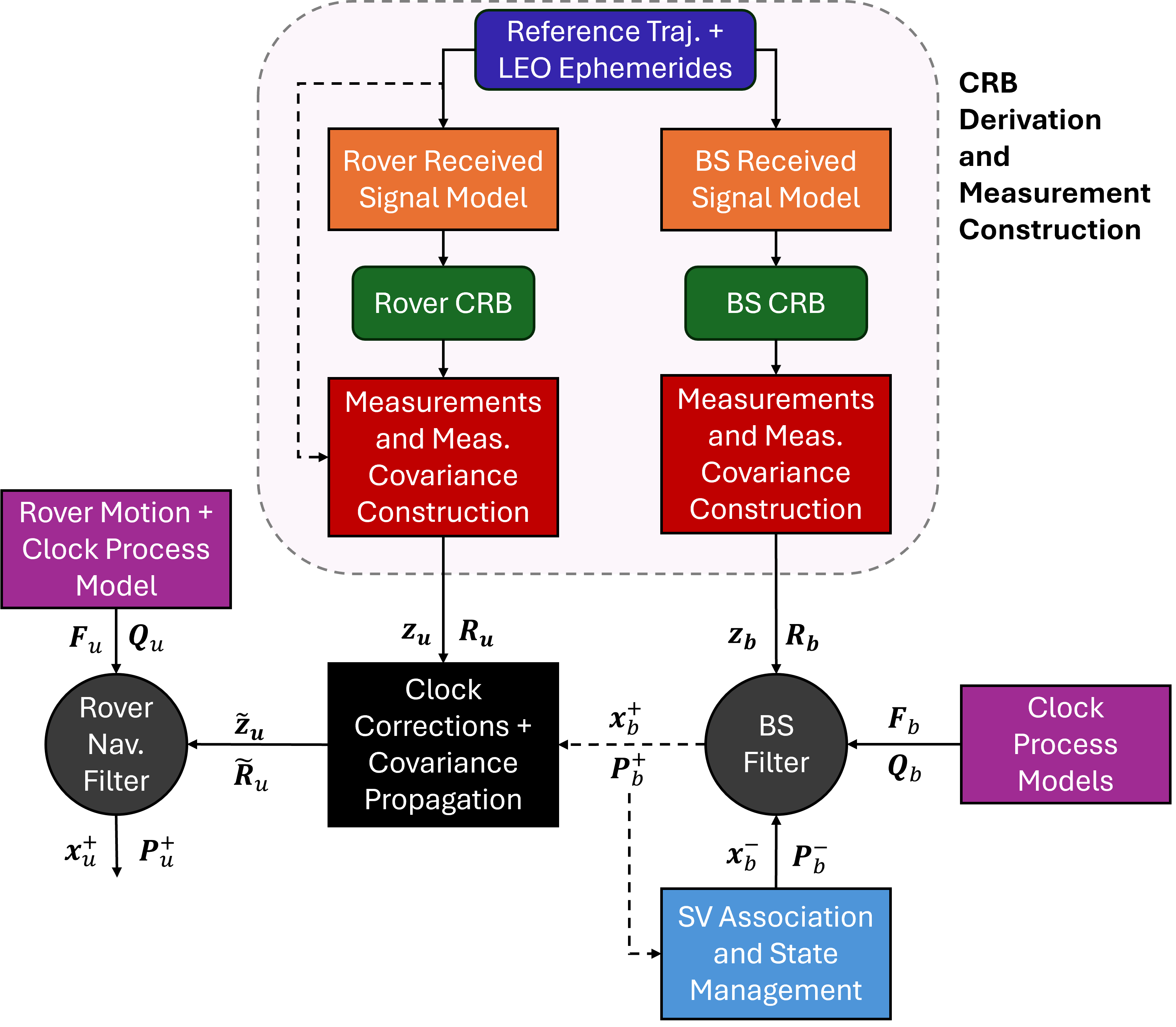}
    \caption{Block diagram of the proposed method.}
    \label{fig:system_overview}
\end{figure}

Fig.~\ref{fig:system_overview} summarizes the proposed base-station--aided differential navigation architecture.
The methodology couples two estimators that operate on the same set of LEO downlink links at each epoch: (i) a BS-side satellite-clock filter that tracks the per-satellite clock bias and drift states over the time-varying visibility set, and (ii) a rover-side navigation filter that estimates the rover motion and receiver clock.
A key design choice is to \emph{avoid} augmenting the rover state with satellite clock states; instead, satellite clock effects are either absorbed as additional measurement uncertainty (rover-only operation) or removed via BS-provided measurement-domain corrections (BS-aided operation).

\subsection{Base Station Clock-Tracking Filter}\label{bs_est}
This section presents the BS clock estimator that tracks the satellite clock bias/drift states required to form differential corrections. 
We assume that, at each epoch, the BS can obtain delay and Doppler observations for each visible satellite using an \emph{efficient} channel-parameter estimator operating on the received communication waveform. 
Accordingly, the BS measurement noise statistics are modeled using the per-link uncertainty derived from the corresponding channel-domain Fisher information bounds established in Section~\ref{sec:crlb}, and are treated as known inputs to the BS filter. In the following subsections, we specify the BS clock-tracking estimator by: (i) establishing the visibility management and state registration mechanism under time-varying satellite geometry; (ii) defining the time-varying state vector, the process model, and its process-noise covariance; (iii) constructing geometry-free measurements; (iv) deriving the corresponding observation model; and (v) specifying the measurement-noise covariance.

\subsubsection{Satellite Visibility Management and State Registration}
\label{subsec:sat_visibility}
The rapid motion of LEO satellites causes frequent changes in the usable visibility set $\mathcal{I}_k$ of size $m_k$ at epoch $k$. To prevent state-to-measurement misalignment and ensure consistent base-station corrections, the stacked $2m_k$ delay/Doppler measurements and per-satellite clock states must be explicitly aligned across consecutive epochs. By comparing current satellite labels to the previous epoch, we partition the visibility sets into persistent ($\mathcal{I}^{\mathrm{keep}}_k = \mathcal{I}_{k-1} \cap \mathcal{I}_k$), appearing ($\mathcal{I}^{\mathrm{new}}_k = \mathcal{I}_k \setminus \mathcal{I}_{k-1}$), and disappearing ($\mathcal{I}^{\mathrm{lost}}_k = \mathcal{I}_{k-1} \setminus \mathcal{I}_k$) satellites. 

Based on this partition, a sequential state registration operation transforms the previous BS posterior $(\mathbf{x}_{b,k-1},\mathbf{P}_{b,k-1})$ and its corresponding information matrix $\mathbf{J}_{b,k-1}$ to match the new epoch's ordering. First, entries for lost satellites in $\mathcal{I}^{\mathrm{lost}}_k$ are marginalized; they are cropped directly from the state vector and covariance matrix, while the corresponding information matrix $\mathbf{J}_{b,k-1} = \mathbf{P}_{b,k-1}^{-1}$ is marginalized via the Schur complement to preserve the exact marginal posterior of the retained satellites. Next, the mean, covariance, and information components of the kept satellites in $\mathcal{I}^{\mathrm{keep}}_k$ are permuted to match the new ordered index of $\mathcal{I}_k$. Finally, new clock states for appearing satellites in $\mathcal{I}^{\mathrm{new}}_k$ are initialized and appended with zero cross-covariance. This is achieved by augmenting the covariance matrix with a tuned prior block $\mathbf{P}_0$ and appending the corresponding inverse block $\mathbf{P}_0^{-1}$ to the information matrix, such that
\begin{equation}\label{eq:new_sat_prior}
\mathbb{E}\!\left(\begin{bmatrix}b_{i,k}\\ d_{i,k}\end{bmatrix}\right) = \begin{bmatrix}0\\0\end{bmatrix}, \qquad \mathrm{Cov}\!\left(\begin{bmatrix}b_{i,k}\\ d_{i,k}\end{bmatrix}\right) = \mathbf{P}_0.
\end{equation}
The choice of $\mathbf{P}_0$ is critical to prevent the filter from becoming overconfident upon satellite entry, which would otherwise yield large consistency violations.

\subsubsection{States and State Transition Model}
The BS clock-tracking state at epoch $k$ stacks the clock states of the visible satellites
\begin{equation}\label{eq:state_bs}
\mathbf{x}_{b,k}=
\begin{bmatrix}
b_{1,k}\\
d_{1,k}\\
\vdots\\
b_{m_k,k}\\
d_{m_k,k}
\end{bmatrix}
=
\begin{bmatrix}
\mathbf{c}_{1,k}\\
\vdots\\
\mathbf{c}_{m_k,k}
\end{bmatrix}
\in\mathbb{R}^{2m_k}.
\end{equation}

The clock dynamics are modeled using a standard discrete-time \emph{integrated Gauss--Markov} drift model, detailed in Section~\ref{subsec:sys_clock_model}. This model is applied to each satellite clock independently. Over the epoch interval $\Delta t$, the $2 \times 2$ state transition matrix for the $i$-th satellite is defined as
\begin{equation}\label{eq:F_clk}
\mathbf{F}^{\text{clk}}_{i} =
\begin{bmatrix}
1 & \Delta t \\
0 & \phi
\end{bmatrix},
\end{equation}
where $\phi$ is the Gauss--Markov correlation coefficient. Let $\mathbf{F}_{b,k}$ denote the global block-diagonal state transition matrix assembled for all active satellites
\begin{equation}\label{eq:F_bs_global}
\mathbf{F}_{b,k} = \mathrm{blkdiag}\big(\mathbf{F}^{\text{clk}}_{1}, \dots, \mathbf{F}^{\text{clk}}_{m_k}\big).
\end{equation}
The resulting recursion is
\begin{equation}\label{eq:bs_state_recursion}
\mathbf{x}_{b,k+1}=\mathbf{F}_{b,k}\,\mathbf{x}_{b,k}+\mathbf{w}_{b,k},
\end{equation}
where $\mathbf{w}_{b,k}$ is the stacked process noise vector with covariance $\mathbf{Q}_{b,k}$.

\subsubsection{Process Covariance Matrix}
Under $\mathbf{F}_{b,k}$, each $(b,d)$ pair admits a $2\times 2$ discrete-time process noise covariance that captures (i) white time-error perturbations on $b$, and (ii) Gauss--Markov driving noise on $d$. Let $\mathbf{Q}^{\text{clk}}_{i}$ denote the corresponding $2\times 2$ covariance for the $i$-th satellite's clock, parameterized by a drift-driving standard deviation $\sigma_{y,i}$, a white time-noise standard deviation $\sigma_{b,i}$, and the epoch interval $\Delta t$. Based on a first-order discrete integration where the drift process noise is integrated into the bias state over $\Delta t$, the exact expression for this matrix is given by
\begin{equation}\label{eq:Q_clk}
\mathbf{Q}^{\text{clk}}_{i} = 
\begin{bmatrix}
\sigma_{b,i}^2 & 0 \\
0 & \sigma_{y,i}^2
\end{bmatrix}.
\end{equation}
Then, the BS process covariance is assembled as a block diagonal matrix
\begin{equation}\label{eq:Qbs_block}
\mathbf{Q}_{b,k}=\mathrm{blkdiag}\Big(
\mathbf{Q}^{\text{clk}}_{1},
\dots,
\mathbf{Q}^{\text{clk}}_{m_k}
\Big).
\end{equation}
This formation reflects the assumption that distinct satellite clocks evolve independently over the considered time scales.

\subsubsection{Measurements}
Since the BS position/velocity are known, it is convenient to remove the purely geometric components from the BS delay/Doppler, defining per-satellite ``clock-only" observations
\begin{align}
z^{\tau}_{b,i,k} &= \tau_{b,i,k}-\frac{\|\mathbf{p}_{i,k}-\mathbf{p}_b\|}{c}
= -b_{i,k} + \eta^\tau_{b,i,k},\label{eq:bs_tau_resid}\\
z^{\nu}_{b,i,k} &= \nu_{b,i,k}+\frac{\mathbf{u}_{b,i,k}^\top\mathbf{v}_{i,k}}{c}
= -d_{i,k} + \eta^\nu_{b,i,k},\label{eq:bs_nu_resid}
\end{align}
where $\eta^\tau_{b,i,k}$ and $\eta^\nu_{b,i,k}$ represent the residual observation noise after geometric removal; their statistics are described in Section~\ref{R}.

Stacking the measurements in an interleaved delay/Doppler form yields
\begin{equation}\label{eq:zbs_stack}
\mathbf{z}_{b,k}=
\big[z^{\tau}_{b,1,k},\,z^{\nu}_{b,1,k},\,\dots,\,z^{\tau}_{b,m_k,k},\,z^{\nu}_{b,m_k,k}\big]^\top\in\mathbb{R}^{2m_k}.
\end{equation}

\subsubsection{Observation Model}
With the ``clock-only" observation construction in \eqref{eq:bs_tau_resid}--\eqref{eq:bs_nu_resid}, the BS observation model is linear in $\mathbf{x}_{b,k}$
\begin{equation}\label{eq:bs_obs_linear}
\mathbf{z}_{b,k}=\mathbf{T}_{b,k}\,\mathbf{x}_{b,k}+\boldsymbol{\eta}_{b,k},
\end{equation}
where $\boldsymbol{\eta}_{b,k}$ stacks $\eta^\tau_{b,i,k}$ and $\eta^\nu_{b,i,k}$ in the same interleaved order as \eqref{eq:zbs_stack}. 

The Jacobian $\mathbf{T}_{b,k}=\partial\mathbf{z}_{b,k}/\partial\mathbf{x}_{b,k} \in \mathbb{R}^{2m_k \times 2m_k}$ is a full-rank matrix with non-zero entries corresponding to the satellite clocks: $\partial z^{\tau}_{b,i,k}/\partial b_{i,k}=-1$ and $\partial z^{\nu}_{b,i,k}/\partial d_{i,k}=-1$. Consequently, the global Jacobian is simply a negative identity matrix
\begin{equation}\label{eq:T_bs}
    \mathbf{T}_{b,k} = -\mathbf{I}_{2m_k}.
\end{equation}

\subsubsection{Measurement Noise Covariance}\label{R}
The BS measurement covariance $\mathbf{R}_{b,k}$ is constructed by aggregating the per-link delay and Doppler uncertainties. Specifically, the $2 \times 2$ measurement noise covariance for link $i$ at epoch $k$, denoted as $\mathbf{R}_{b,i,k}$, is determined by the channel-domain information bounds derived in Section~\ref{sec:crlb}. 
Assuming independent receiver noise across frequency-multiplexed satellite links, $\mathbf{R}_{b,k}$ is block diagonal
\begin{equation}\label{eq:Rbs_block}
\mathbf{R}_{b,k}=
\mathrm{blkdiag}\Big(
\mathbf{R}_{b,1,k},\,
\mathbf{R}_{b,2,k},\,
\dots,\,
\mathbf{R}_{b,m_k,k}
\Big),
\end{equation}

Finally, the BS state is estimated recursively using a Kalman filter driven by \eqref{eq:bs_state_recursion} and \eqref{eq:bs_obs_linear}, with $\mathbf{Q}_{b,k}$ and $\mathbf{R}_{b,k}$ constructed as described above. 
The resulting posterior $(\hat{\mathbf{x}}_{b,k},\mathbf{P}_{b,k})$ provides both the per-satellite clock corrections and their uncertainty, which are subsequently propagated to the rover measurement domain in Section~\ref{subsec:bs_aided}.

\subsection{Rover Kinematic Navigation Filter}\label{ms_est}
This section presents the rover navigation filter that jointly tracks rover kinematics and receiver clock bias/drift using delay and Doppler measurements from the satellites visible at epoch $k$. 
In contrast to the BS estimator, the rover filter maintains a \emph{compact} state and does not augment per-satellite clock states. 

\subsubsection{States and State Transition Model}
The rover state $\mathbf{x}_{u,k}$ is defined as
\begin{equation}\label{eq:state_rover}
\mathbf{x}_{u,k}=
\begin{bmatrix}
\mathbf{p}_{u,k}\\
\mathbf{v}_{u,k}\\
b_{u,k}\\
d_{u,k}
\end{bmatrix}
\in\mathbb{R}^{8}.
\end{equation}

The rover dynamics follow a constant-velocity (CV) kinematic model with discrete-time clock evolution consistent with the integrated Gauss--Markov drift model. 
The state recursion is
\begin{equation}\label{eq:rover_state_recursion}
\mathbf{x}_{u,k+1}=\mathbf{F}_{u}\,\mathbf{x}_{u,k}+\mathbf{w}_{u,k},
\end{equation}
with a block-structured transition matrix
\begin{equation}\label{eq:F_rover}
\mathbf{F}_{u}=
\begin{bmatrix}
\mathbf{I}_3 & \Delta t\,\mathbf{I}_3 & \mathbf{0}_{3\times 1} & \mathbf{0}_{3\times 1}\\
\mathbf{0}_{3\times 3} & \mathbf{I}_3 & \mathbf{0}_{3\times 1} & \mathbf{0}_{3\times 1}\\
\mathbf{0}_{1\times 3} & \mathbf{0}_{1\times 3} & 1 & \Delta t\\
\mathbf{0}_{1\times 3} & \mathbf{0}_{1\times 3} & 0 & \phi_u
\end{bmatrix},
\end{equation}
where $\phi_u\in(0,1)$ is the Gauss--Markov coefficient of the rover clock drift process and $\mathbf{w}_{u,k}$ is zero-mean process noise with covariance $\mathbf{Q}_{u}$.

\subsubsection{Process Covariance Matrix}
The rover process covariance is chosen to reflect kinematic maneuvering uncertainty and receiver clock driving noise. For the constant-velocity (CV) model, the position and velocity block corresponds to discretized continuous-time white acceleration noise. Let $\mathbf{\Sigma}_a = \mathrm{diag}(\sigma_{a,x}^2, \sigma_{a,y}^2, \sigma_{a,z}^2)$ denote the diagonal matrix of the acceleration variances (in $\mathrm{m^2/s^4}$) along each Cartesian axis. The exact $6 \times 6$ discrete-time process noise covariance for the kinematic states over the interval $\Delta t$ is given by
\begin{equation}\label{eq:Q_cv}
\mathbf{Q}^{\text{cv}} = 
\begin{bmatrix}
\frac{\Delta t^3}{3} \mathbf{\Sigma}_a & \frac{\Delta t^2}{2} \mathbf{\Sigma}_a \\
\frac{\Delta t^2}{2} \mathbf{\Sigma}_a & \Delta t \mathbf{\Sigma}_a
\end{bmatrix}.
\end{equation}

For the receiver clock, the process noise $\mathbf{Q}^{\text{clk}}_{u}$ follows the same two-noise structure established for the satellite clocks in \eqref{eq:Q_clk}, parameterized by the rover's white time-noise standard deviation and drift-driving standard deviation. Accordingly, the full $8 \times 8$ rover process covariance matrix is assembled as a block-diagonal matrix
\begin{equation}\label{eq:Qu_block}
\mathbf{Q}_u = \mathrm{blkdiag}\big(\mathbf{Q}^{\text{cv}},\, \mathbf{Q}^{\text{clk}}_{u}\big).
\end{equation}

\subsubsection{Measurements}
Let $\mathcal{I}_k$ denote the set of satellites that provide usable rover measurements at epoch $k$.
For each $i\in\mathcal{I}_k$, the rover forms a delay and Doppler observation pair $(\tau_{u,i,k},\nu_{u,i,k})$.
Stacking in the same interleaved form used throughout the paper gives
\begin{equation}\label{eq:zrover_stack}
\mathbf{z}_{u,k}=
\big[\tau_{u,1,k},\,\nu_{u,1,k},\,\dots,\,\tau_{u,m_k,k},\,\nu_{u,m_k,k}\big]^\top\in\mathbb{R}^{2m_k}.
\end{equation}

\subsubsection{Observation Model}
The Jacobian $\mathbf{T}_{u,i,k}=\partial\mathbf{z}_{u,i,k}/\partial\mathbf{x}_{u,k} \in \mathbb{R}^{2 \times 8}$ yields
\begin{equation}\label{eq:T_rover}
\mathbf{T}_{u,i,k}=
\begin{bmatrix}
\displaystyle \frac{\partial \tau_{u,i,k}}{\partial \mathbf{p}_{u,k}} &
\displaystyle \frac{\partial \tau_{u,i,k}}{\partial \mathbf{v}_{u,k}} &
\displaystyle \frac{\partial \tau_{u,i,k}}{\partial b_{u,k}} &
\displaystyle \frac{\partial \tau_{u,i,k}}{\partial d_{u,k}}
\\[8pt]
\displaystyle \frac{\partial \nu_{u,i,k}}{\partial \mathbf{p}_{u,k}} &
\displaystyle \frac{\partial \nu_{u,i,k}}{\partial \mathbf{v}_{u,k}} &
\displaystyle \frac{\partial \nu_{u,i,k}}{\partial b_{u,k}} &
\displaystyle \frac{\partial \nu_{u,i,k}}{\partial d_{u,k}}
\end{bmatrix}.
\end{equation}
The individual terms take the standard closed form
\begin{equation}
\begin{aligned}
\frac{\partial \tau_{u,i,k}}{\partial \mathbf{p}_{u,k}} &= -\frac{1}{c}\mathbf{u}_{u,i,k}^\top, &
\frac{\partial \tau_{u,i,k}}{\partial \mathbf{v}_{u,k}} &= \mathbf{0}_{1\times 3}, &\\
\frac{\partial \tau_{u,i,k}}{\partial b_{u,k}} &= 1, &
\frac{\partial \tau_{u,i,k}}{\partial d_{u,k}} &= 0, \label{eq:dtau_terms}\\[4pt]
\end{aligned}
\end{equation}
and the state sensitivity to Doppler is
\begin{align} \label{eq:dnu_terms_simple}
\frac{\partial \nu_{u,i,k}}{\partial \mathbf{p}_{u,k}} &= \frac{1}{c}(\mathbf{v}_{i,k}-\mathbf{v}_{u,k})^\top\left(\frac{\mathbf{I}_3-\mathbf{u}_{u,i,k}\mathbf{u}_{u,i,k}^\top}{\|\mathbf{p}_{i,k}-\mathbf{p}_{u,k}\|}\right), \\
\frac{\partial \nu_{u,i,k}}{\partial \mathbf{v}_{u,k}} &= \frac{1}{c}\mathbf{u}_{u,i,k}^\top, \quad
\frac{\partial \nu_{u,i,k}}{\partial b_{u,k}} = 0, \quad
\frac{\partial \nu_{u,i,k}}{\partial d_{u,k}} = 1.
\end{align}
Stacking \eqref{eq:T_rover} over all $i\in\mathcal{I}_k$ yields the rover measurement Jacobian $\mathbf{T}_{u,k}\in\mathbb{R}^{2m_k\times 8}$ and the nonlinear measurement model
\begin{equation}\label{eq:rover_obs_model}
\mathbf{z}_{u,k}=\mathbf{h}_u(\mathbf{x}_{u,k})+\boldsymbol{\eta}_{u,k},
\end{equation}
which is implemented via an EKF update using $\mathbf{T}_{u,k}$.

\subsubsection{Measurement Noise Covariance}
Similar to the BS, the baseline rover measurement covariance is constructed using the per-link delay and Doppler uncertainties established via the information-theoretic bounds in Section~\ref{sec:crlb}. Denoting the $2 \times 2$ covariance block for the $i$-th visible satellite as $\mathbf{R}_{u,i,k}$, and applying the same independence assumptions across frequency-multiplexed links, the rover measurement covariance is block diagonal
\begin{equation}\label{eq:Ru_block}
\mathbf{R}_{u,k}= \mathrm{blkdiag}\Big(\mathbf{R}_{u,1,k},\,
\dots,\, \mathbf{R}_{u,m_k,k} \Big).
\end{equation}
This baseline covariance matrix would be inflated depending on the operation mode, be it \emph{rover-only} or \emph{BS-aided}, as detailed in the next section.

\subsection{Differential Assistance \& Uncertainty Propagation}
\subsubsection{Rover-only Operation}\label{subsec:rover_only}
In rover-only operation, the rover navigation filter processes delay/Doppler observations without any external clock assistance. As the rover state does not include satellite clock states, the per-link satellite clock terms act as unmodeled components in the delay/Doppler observations. Therefore, rover-only operation treats these terms as an additional source of uncertainty in the measurement domain. Practically, this is handled by inflating the measurement covariance via
\begin{equation}\label{eq:R_rover_only}
\mathbf{R}^{(\mathrm{ro})}_{k}
=
\mathbf{R}_{u,k}
+
\mathbf{R}^{\mathrm{sat}}_{k},
\end{equation}
where $\mathbf{R}^{\mathrm{sat}}_{k}$ represents the effective covariance induced by satellite clock bias/drift.
It is obtained by propagating the uncertainty of the unmodeled per-satellite clock bias/drift into the delay/Doppler domain.
Let the true, unmodeled satellite clock state vector at epoch $k$ be strictly defined by $\mathbf{x}_{b,k}$ as formulated in Eq.~\eqref{eq:state_bs}. We denote its open-loop covariance as $\mathbf{P}^{\mathrm{clk}}_{k}\in\mathbb{R}^{2m_k\times 2m_k}$, which quantifies the uncertainty of the unmodeled satellite clock bias/drift. This covariance matrix can be tracked through a standalone clock-dynamics covariance propagation.
For each visible satellite $i\in\mathcal{I}_k$, we consider the two-state clock vector $\mathbf{c}_{i,k}$, evolving according to the Gauss--Markov drift model with integrated bias,
\begin{equation}
\mathbf{c}_{i,k+1} = \mathbf{F}^{\text{clk}}_{i}\,\mathbf{c}_{i,k} + \mathbf{w}_{i,k},
\end{equation}
where $\mathbf{w}_{i,k}$ is zero-mean process noise with covariance $\mathbf{Q}^{\mathrm{clk}}_i$ as defined in \eqref{eq:Q_clk}.
Accordingly, the per-satellite covariance is propagated as
\begin{equation}\label{eq:Pclk_prop}
\mathbf{P}_{i,k+1} = \mathbf{F}^{\text{clk}}_{i}\,\mathbf{P}_{i,k}\,(\mathbf{F}^{\text{clk}}_{i})^\top + \mathbf{Q}^{\text{clk}}_{i}.
\end{equation}

Assuming independent satellite clocks, the stacked open-loop satellite clock covariance over the visible set is block diagonal,
\begin{equation}\label{eq:Psv_stack}
\mathbf{P}^{\mathrm{clk}}_{k}
=
\mathrm{blkdiag}\big(\mathbf{P}_{1,k},\ldots,\mathbf{P}_{m_k,k}\big)\in\mathbb{R}^{2m_k\times 2m_k}.
\end{equation}
When satellites enter or exit the visibility set, this stacked covariance is remapped as detailed in Section~\ref{subsec:sat_visibility}.

Finally, $\mathbf{P}^{\mathrm{clk}}_{k}$ is projected into the rover measurement domain. Because the unmodeled satellite clocks affect the rover delay and Doppler exactly as they affect the BS measurements, the linear clock-to-measurement mapping is governed by the exact same sparse matrix $\mathbf{T}_{b,k}$ defined in \eqref{eq:T_bs}. The effective measurement disturbance is
\begin{equation}
\delta\mathbf{z}^{\mathrm{clk}}_{u,k} = \mathbf{T}_{b,k}\,\mathbf{x}_{b,k} = -\mathbf{x}_{b,k},
\end{equation}
yielding the inflated measurement covariance
\begin{equation}\label{eq:Rsv_fromPsv}
\mathbf{R}^{\mathrm{sat}}_{k}
=
\mathbf{T}_{b,k}\,
\mathbf{P}^{\mathrm{clk}}_{k}\,
\mathbf{T}_{b,k}^\top = 
\mathbf{P}^{\mathrm{clk}}_{k}.
\end{equation}

The rover-only filter update then uses $\mathbf{R}^{(\mathrm{ro})}_{k}$, so that the filter remains statistically consistent even when satellite clock effects are not explicitly estimated.
This mode provides a baseline solution that preserves a compact rover state and enables fair comparisons to the BS-aided solution.

\subsubsection{BS-Aided Differential Operation}
\label{subsec:bs_aided}
In the BS-aided mode, the rover leverages assistance from a nearby base station that continuously tracks the satellite clock states using its own delay/Doppler observations and known kinematics.
At each epoch $k$, the BS filter produces its posterior state estimate of the active satellite clocks $\widehat{\mathbf{x}}_{b,k}$, together with the associated estimation covariance $\mathbf{P}_{b,k}$ defined over the current visibility set $\mathcal{I}_k$.

\paragraph*{Measurement-domain correction}
The rover applies the BS satellite clock estimates directly in the measurement domain.
Using the same interleaved stacking of $\mathbf{z}_{u,k}$, the correction contribution is
\begin{equation}
\widehat{\delta\mathbf{z}}_{b,k}
=
\mathbf{T}_{b,k}\,\widehat{\mathbf{x}}_{b,k}
=
-\widehat{\mathbf{x}}_{b,k}.
\end{equation}
The BS-aided corrected rover measurement vector is then
\begin{equation}\label{eq:z_corr}
\widetilde{\mathbf{z}}_{u,k} = \mathbf{z}_{u,k} - \widehat{\delta\mathbf{z}}_{b,k}.
\end{equation}

\paragraph*{Uncertainty propagation into the rover measurement covariance}
Because the BS state estimate $\widehat{\mathbf{x}}_{b,k}$ is itself uncertain, the correction in Eq.~\eqref{eq:z_corr} introduces an additional error term.
Assuming the BS satellite-clock estimation error is independent of the rover's channel/thermal measurement errors, the corrected-measurement covariance is additively inflated. It is important to note that because $\widehat{\mathbf{x}}_{b,k}$ is generated by a recursive dynamic filter, its estimation errors are inherently time-correlated. However, because the base station operates from a known fixed location with a tightly disciplined time reference, its posterior estimation uncertainty is very small. Consequently, the magnitude of this time-correlated error is heavily dominated by the rover's local, uncorrelated measurement noise $\mathbf{R}_{u,k}$. Therefore, we adopt a practical white-noise approximation by directly injecting the epoch-wise BS posterior covariance into the rover measurement covariance
\begin{equation}\label{eq:R_corr}
\widetilde{\mathbf{R}}_{u,k}
=
\mathbf{R}_{u,k}
+
\mathbf{R}^{\mathrm{corr}}_{b,k},
\end{equation}
where the correction-induced covariance is obtained by linear propagation via the shared Jacobian,
\begin{equation}\label{eq:R_sat_BS}
\mathbf{R}^{\mathrm{corr}}_{b,k}
=
\mathbf{T}_{b,k}\,
\mathbf{P}_{b,k}\,
\mathbf{T}_{b,k}^\top= 
\mathbf{P}_{b,k}.
\end{equation}

\section{Cram\'er--Rao Lower Bound Derivation}
\label{sec:crlb}
\subsection{Channel-Domain Fisher Information Matrix}\label{subsec:channel_fim}
Under the signal model in \eqref{eq:rx_signal}--\eqref{eq:channel_model}, the information about the unknown parameters is quantified through the Fisher information matrix (FIM). For a given link $(r,i,k)$, we collect the channel-domain parameters as
\begin{equation}
\boldsymbol{\theta}_{r,i,k}=
\begin{bmatrix}
\Re\{\alpha_{r,i,k}\} & \Im\{\alpha_{r,i,k}\} & \tau_{r,i,k} & \nu_{r,i,k}
\end{bmatrix}^\top\in\mathbb{R}^{4},
\end{equation}
where $\Re\{\cdot\}$ and $\Im\{\cdot\}$ denote, respectively, the real and imaginary parts of a complex quantity.

Let $\zeta^{n,l}_{r,k}$ denote the noise-free received signal defined in \eqref{eq:rx_signal}. The per-link FIM for $\boldsymbol{\theta}_{r,i,k}$ is then
\begin{equation}\label{eq:fim_channel}
\boldsymbol{J}^{\text{ch}}_{r,i,k}=\frac{2}{\sigma^{2}_r}\sum_{l=0}^{L-1}\sum_{n=0}^{N-1}\Re\left\{\left(\frac{\partial {\zeta}^{n,l}_{r,k}}{\partial\boldsymbol{\theta}_{r,i,k}}\right)^{H}\left(\frac{\partial {\zeta}^{n,l}_{r,k}}{\partial\boldsymbol{\theta}_{r,i,k}}\right)\right\}, 
\end{equation}

Since the gain components $\Re\{\alpha_{r,i,k}\},\Im\{\alpha_{r,i,k}\}$ are nuisance parameters for positioning and timing, we work with the equivalent Fisher information matrix (EFIM) for the delay--Doppler pair.
Let $\boldsymbol{\theta}_{r,i,k}=\big[\boldsymbol{\alpha}_{r,i,k}^\top\;\;\boldsymbol{\beta}_{r,i,k}^\top\big]^\top$, where
$\boldsymbol{\alpha}_{r,i,k}=\big[\Re\{\alpha_{r,i,k}\}\;\;\Im\{\alpha_{r,i,k}\}\big]^\top$ and
$\boldsymbol{\beta}_{r,i,k}=\big[\tau_{r,i,k}\;\;\nu_{r,i,k}\big]^\top$.
Accordingly, we partition the channel-parameter FIM as
\begin{equation}\label{eq:fim_partition}
\mathbf{J}^{\mathrm{ch}}_{r,i,k}=
\begin{bmatrix}
\mathbf{J}_{\alpha\alpha} & \mathbf{J}_{\alpha\beta}\\
\mathbf{J}_{\beta\alpha}  & \mathbf{J}_{\beta\beta}
\end{bmatrix}.
\end{equation}

The EFIM for the delay--Doppler pair $\boldsymbol{\beta}_{r,i,k}$ is then given by the Schur complement
\begin{equation}\label{eq:efim_taunu}
\mathbf{J}^{\beta}_{r,i,k}
=
\mathbf{J}_{\beta\beta}
-\mathbf{J}_{\beta\alpha}\mathbf{J}_{\alpha\alpha}^{-1}\mathbf{J}_{\alpha\beta}.
\end{equation}

Finally, the corresponding per-link CRLB covariance for $\boldsymbol{\beta}_{r,i,k}$ is
\begin{equation}\label{eq:crlb_taunu}
\mathbf{C}^{\beta}_{r,i,k} = \left(\mathbf{J}^{\beta}_{r,i,k}\right)^{-1}\in\mathbb{R}^{2\times 2}.
\end{equation}

\subsection{State-Domain Fisher Information Matrix}\label{subsec:state_fim}
Next, we map channel-domain information to the state domain using the Jacobian matrix $\mathbf{T}_{r,i,k}$ defined in \eqref{eq:T_bs} and \eqref{eq:T_rover} for the BS and the UE, respectively.


Under the standard independence assumption across satellites enabled by the frequency-multiplexed pilot structure, the measurement information contributions add across links. Hence, the instantaneous (non-recursive) state-domain measurement FIM at epoch $k$ is
\begin{equation}\label{eq:state_fim_sum}
\mathbf{J}^{\text{meas}}_{r,k}
=
\sum_{i\in\mathcal{I}_k}
\mathbf{T}_{r,i,k}^\top\,
\mathbf{J}^{\beta}_{r,i,k}\,
\mathbf{T}_{r,i,k},
\end{equation}
which captures how link geometry, SNR, and pilot resources (through $\mathbf{J}^{\beta}_{r,i,k}$) translate into information about the state $\mathbf{x}_{r,k}$. 

Subsequently, the instantaneous state-domain CRLB at epoch $k$ is obtained by inverting the corresponding measurement FIM when it is nonsingular
\begin{equation}\label{eq:snapshot_crlb}
\mathbf{C}^{\mathrm{inst}}_{r,k}=\left(\mathbf{J}^{\mathrm{meas}}_{r,k}\right)^{-1},
\end{equation}
which quantifies the best achievable estimation accuracy from the measurements at epoch $k$ alone.

\subsection{Recursive Bayesian Cramér-Rao Bound (RBCRB)}
\label{sec:rbcrb}
\subsubsection{CRB Recursion}
To evaluate the fundamental performance limits of the proposed tracking algorithm, we utilize the recursive Bayesian Cramér-Rao bound (RBCRB). Unlike the snapshot CRB derived in the previous section, the RBCRB accounts for the temporal evolution of the state and the accumulation of information over successive epochs. 

Following the recursive framework established in \cite{tichavsky1998posterior}, the Bayesian information matrix (BIM) $\mathbf{J}_{r,k+1}$ is decomposed into the information contribution from the current measurement and the predicted information from the previous state estimate
\begin{equation}\label{eq:RBCRB_general}
\mathbf{J}_{r,k+1} = \mathbf{J}^{\mathrm{meas}}_{r,k+1} + \mathbf{J}^{\mathrm{prior}}_{r,k+1},
\end{equation}
where $\mathbf{J}^{\mathrm{meas}}_{r,k+1}$ is the measurement information matrix derived in \eqref{eq:state_fim_sum}. The term $\mathbf{J}^{\mathrm{prior}}_{r,k+1}$ represents the information propagated from the posterior at epoch $k$ through the system dynamics. Under the assumption of additive Gaussian process noise, this prior information is computed via the Riccati-like recursion
\begin{equation}\label{eq:prior_schur}
\mathbf{J}^{\mathrm{prior}}_{r,k+1} = \mathbf{D}^{22}_{r,k} - \mathbf{D}^{21}_{r,k} \left( \mathbf{J}_{r,k} + \mathbf{D}^{11}_{r,k} \right)^{-1} \mathbf{D}^{12}_{r,k},
\end{equation}
where the intermediate information blocks related to the dynamics are defined as
\begin{subequations}\label{eq:D_blocks}
\begin{align}
\mathbf{D}^{11}_{r,k} &= \mathbf{F}_{r,k}^\top \mathbf{Q}_{r,k}^{-1} \mathbf{F}_{r,k}, \\
\mathbf{D}^{12}_{r,k} &= -\mathbf{F}_{r,k}^\top \mathbf{Q}_{r,k}^{-1} = (\mathbf{D}^{21}_{r,k})^\top, \\
\mathbf{D}^{22}_{r,k} &= \mathbf{Q}_{r,k}^{-1}.
\end{align}
\end{subequations}


By iteratively updating $\mathbf{J}_{r,k}$ using \eqref{eq:RBCRB_general}--\eqref{eq:D_blocks}, we obtain a recursive lower bound on the estimation error covariance. This provides a benchmark to assess how closely our tracking filter approaches the theoretical limit of information extraction from satellite signals.

\subsubsection*{Effect of BS Assistance on the Theoretical Bounds}
To quantify the fundamental performance limits under BS assistance, the theoretical bounds must account for the estimation uncertainty injected by the measurement-domain corrections. Assuming the base station's clock estimation errors are statistically independent of the rover's local channel and thermal noise, the effective observation-domain CRLB covariance is given by the additive sum of their respective uncertainties
\begin{equation}\label{eq:Reff_bs_aided}
\widetilde{\mathbf{C}}_{u,k}^{\beta}
=
\mathbf{C}^{\beta}_{u,k}
+
\mathbf{T}_{b,k}\,
\mathbf{C}_{b,k}\,
\mathbf{T}_{b,k}^{\top}= 
\mathbf{C}^{\beta}_{u,k}
+
\mathbf{C}_{b,k},
\end{equation}
where $\mathbf{C}^{\beta}_{u,k}=(\mathbf{J}^{\beta}_{u,k})^{-1}$ is the inherent delay/Doppler covariance bound of the rover links, and $\mathbf{C}_{b,k}$ is the theoretical posterior bound of the base station's satellite clock estimates, extracted directly from the BS RBCRB recursion.

To map this effective measurement uncertainty into the kinematic state domain, we evaluate the rover measurement Jacobian $\mathbf{T}_{u,k}$ along the true reference state (e.g., the ground-truth trajectory $\mathbf{x}^{\star}_{u,k}$). Denoting this ideal geometry matrix as $\bar{\mathbf{T}}_{u,k}$, the effective state-domain measurement information matrix for the BS-aided architecture is defined as
\begin{equation}\label{eq:Jmeas_Reff}
\widetilde{\mathbf{J}}^{\mathrm{meas}}_{u,k}
=
\bar{\mathbf{T}}_{u,k}^{\top}\,
\big(\widetilde{\mathbf{C}}_{u,k}^{\beta}\big)^{-1}\,
\bar{\mathbf{T}}_{u,k}.
\end{equation}

Finally, substituting the aided measurement information matrix $\widetilde{\mathbf{J}}^{\mathrm{meas}}_{u,k}$ into the information update step of the RBCRB recursion \eqref{eq:RBCRB_general} yields the theoretical, dynamically propagated tracking bounds for the proposed differential framework. 

\subsubsection{Performance metrics derived from the RBCRB}
From $\mathbf{J}_{u,k}^{-1}$, we report the following rover bounds

\begin{subequations}\label{eq:peb_RBCRB}
\begin{align}
\mathrm{PEB}_k &= \sqrt{\mathrm{tr}\!\left(\mathbf{J}_{u,k}^{-1}(1{:}3,1{:}3)\right)},\\
\mathrm{VEB}_k &= \sqrt{\mathrm{tr}\!\left(\mathbf{J}_{u,k}^{-1}(4{:}6,4{:}6)\right)},\\
\mathrm{CBEB}_k &= \sqrt{\mathbf{J}_{u,k}^{-1}(7,7)},\\
\mathrm{CDEB}_k &= \sqrt{\mathbf{J}_{u,k}^{-1}(8,8)}.
\end{align}
\end{subequations}

For the BS estimator, clock-only bounds are obtained from $\mathbf{J}_{b,k}^{-1}$ by selecting the corresponding state entries.

\section{Simulation Setup} \label{sec:road test setup}
\subsection{Reference Trajectory Data Collection}
\label{subsec:ref_traj}
\begin{figure}
    \centering
    \includegraphics[width=\linewidth]{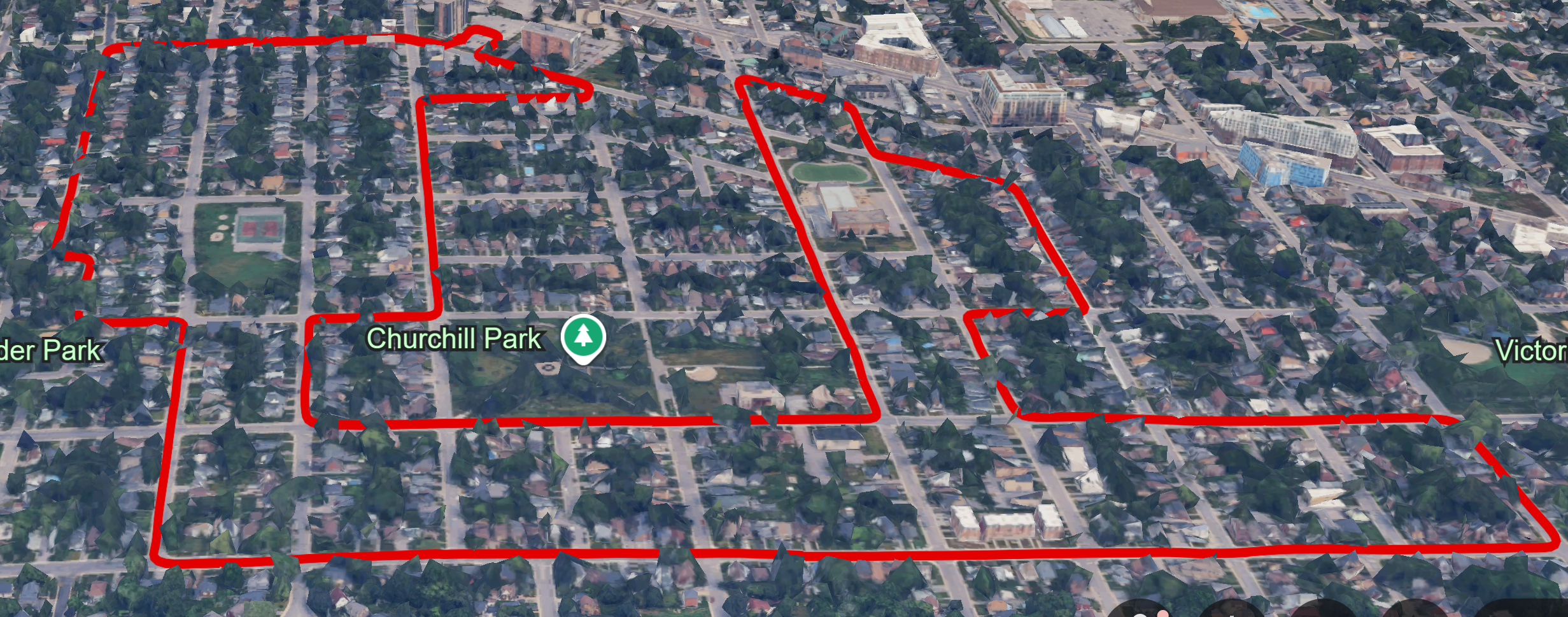}
    \caption{\texttt{Urban03} trajectory collected in downtown Kingston, Canada.}
    \label{fig:traj}
\end{figure}

To evaluate the proposed framework under realistic user dynamics, we employ a road-test reference trajectory collected using a high-precision positioning system mounted on a passenger vehicle. The reference platform integrates a geodetic-grade GNSS receiver (NovAtel PwrPak7) with a tactical-grade inertial measurement unit (KVH1750), providing a tightly coupled GNSS/INS solution that serves as ground truth for the rover position and velocity. The experiment was conducted in Kingston, Ontario, and the recorded drive lasted approximately 19 minutes. In this work, we use the \texttt{Urban03} trajectory shown in Fig.~\ref{fig:traj} from the \texttt{NavINST} dataset, which is publicly available (see \cite{de2025navinst}). 

Using measured road-test data is preferred over synthetic motion profiles because it naturally captures driving behaviors that are difficult to model faithfully, including stop-and-go patterns, acceleration and braking transients, turning maneuvers with realistic curvature, and speed variations induced by road grade and terrain changes. These effects yield a representative velocity and acceleration spectrum for stress-testing both the rover motion model and the clock/measurement fusion strategy in an urban environment.

\subsection{LEO Satellite Simulation}
To emulate a realistic time-varying LEO geometry along the road-test trajectory, we use the Skydel GNSS/LEO simulator as a kinematic propagation engine for satellite motion and visibility. In particular, the simulator is configured to generate the satellite ephemerides corresponding to XONA's \emph{PULSAR} constellation over the duration of the experiment. The reference trajectory is then provided to Skydel so that satellite–receiver line-of-sight conditions are evaluated consistently with the vehicle motion over an elevation mask of $10^\circ$. 

\begin{figure}
        \centering
        \includegraphics[width=\linewidth]{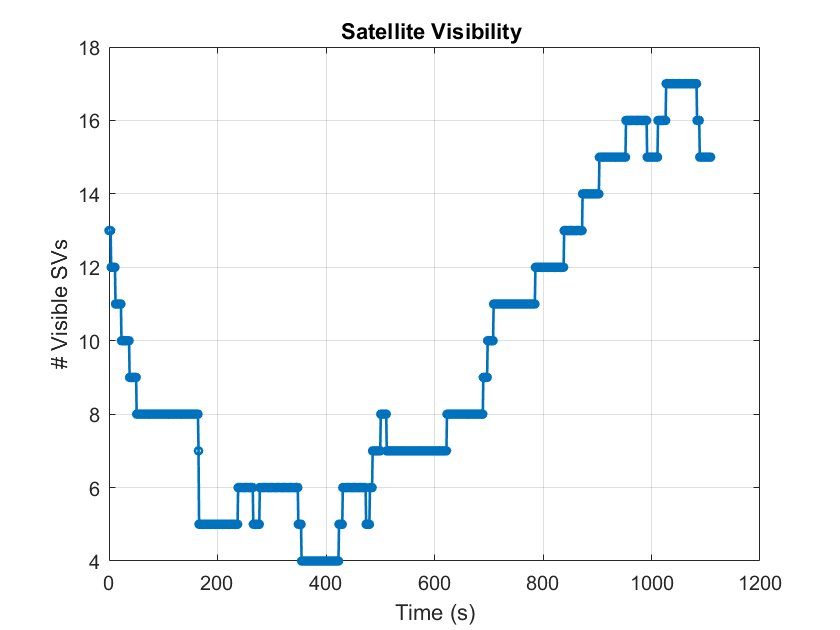}
        \caption{Number of satellites available along the trajectory.}
        \label{fig:sat_visibility}
\end{figure}

To illustrate the temporal variability in satellite availability along the route, we plot the number of visible satellites as a function of time (Fig.~\ref{fig:sat_visibility}). 

\subsection{Clock Error Simulation Model}
\label{subsec:clock-sim}
For the numerical evaluation, synthetic clock errors were generated according to the stochastic dynamic model established in Section~\ref{subsec:sys_clock_model}. To reflect a realistic architectural scenario where the fixed and space segments utilize high-stability references while the user segment relies on less expensive commercial oscillators, the LEO satellites were parameterized as chip-scale atomic clocks (CSACs), whereas the mobile rover was equipped with an oven-controlled crystal oscillator (OCXO). Table~\ref{tab:clock-params} summarizes the specific correlation times and driving noise standard deviations utilized to parameterize the state evolution in Eqs.~\eqref{eq:gm-drift} and \eqref{eq:bias-update} \cite{flood2023formation,s16050682}.

\begin{table}[h]
\centering
\caption{Clock parameters used in the simulation.}
\label{tab:clock-params}
\begin{tabular}{lcc}
\hline
 & \textbf{CSAC (sat)} & \textbf{OCXO (rover)} \\
\hline 
Correlation time $\tau$ [s]      & $300$              & $100$ \\
Frequency noise $\sigma_y$ [s/s] & $1 \times 10^{-10}$ & $3 \times 10^{-9}$ \\
Time noise $\sigma_b$ [s]        & $1 \times 10^{-11}$ & $3 \times 10^{-10}$ \\
\hline
\end{tabular}
\end{table}
\subsection{LEO Signal and Receiver Configuration}
Table \ref{sim} summarizes the simulation parameters, including the satellite signal characteristics and receiver configuration. The receiver configuration, carrier frequency, and OFDM numerology strictly follow the 3GPP TR 38.811 reference scenario for S-band handheld non-terrestrial networks (NTN) \cite{3gpp_tr38821}. However, the simulated satellite antenna gain diverges from the $30~\text{dBi}$ telecom baseline. This modification reflects the specific orbital geometry of the LEO constellation considered in this work, which is modeled after the Xona constellation ($258$ satellites at an altitude of $1,080~\text{km}$). Unlike typical broadband mega-constellations, such as Starlink, which operates over $9,800$ active satellites as of early 2026 and utilizes high-gain, narrow spot beams for localized throughput, a sparser, higher-altitude navigation constellation requires a lower-gain, wide-area beam (e.g., with $10~\text{dBi}$).

\begin{table}[]
  \caption{Simulation parameters of LEO signals and receiver configuration.}
  \label{sim}
  \centering
  \begin{tabularx}{\columnwidth}{@{}l l l@{}}
    \toprule
    \textbf{Block} & \textbf{Parameter}               & \textbf{Value}\\
    \midrule
    LEO Signal     & Carrier freq.\ \(f_c\)             & 2\,GHz\\
                   & Subcarrier spacing \(\Delta f\)    & 60\,kHz\\
                   & \# subcarriers \(N\)               & 720\\
                   & \# OFDM symbols \(L\)              & 252\\
                   & Bandwidth \(B=N\,\Delta f\)        & 43.2\,MHz\\
                   & Cyclic prefix duration \(T_{\rm cp}\)    & 0.07\,$T$\\
                   & Transmit power                     & 54\,dBm\\
                   & Antenna gain                       & 10\,dBi \\
    \midrule
    Receiver       
                   & Filter sampling interval $\Delta t$ & 1\,s \\
                   & BS position $\mathbf{p}_b$     & $\mathbf{p}_u^0$\\
                   & Antenna gain (BS/UT)               & 0\,dBi\\
                   & Noise figure \textit{NF}                & 7\,dB\\
    \bottomrule
  \end{tabularx}
\end{table}
\subsection{Estimator Initialization and Tuning}
\label{subsec:estimator_tuning}
To ensure reproducible filter convergence and prevent numerical instability during satellite transitions, the prior covariance for newly appearing LEO satellites, $\mathbf{P}_0$, must be carefully tuned. As defined in Section \ref{subsec:sat_visibility}, when a satellite $i$ enters the visibility set $\mathcal{I}_k$, its clock bias and drift states are initialized with zero cross-covariance and a diagonal prior covariance matrix:
\begin{equation}
\mathbf{P}_0 = 
\begin{bmatrix} 
\sigma_{b,0}^2 & 0 \\ 
0 & \sigma_{d,0}^2 
\end{bmatrix},
\end{equation}
where the initial standard deviations were set to $\sigma_{b,0} = 1\times10^{-8}$~s for the clock bias and $\sigma_{d,0} = 5\times10^{-10}$~s/s for the clock drift. 

\subsection{Performance Metrics and Baselines}
\label{subsec:metrics_baselines}
The proposed framework is evaluated against a non-differential \textit{rover-only} baseline (Section~\ref{subsec:rover_only}), which treats asynchronous LEO clocks as unmodeled nuisance parameters, and against the fundamental information-theoretic limits established by the snapshot and recursive Bayesian CRBs. System optimality is directly quantified via the theoretical error bounds in \eqref{eq:peb_RBCRB}. Statistical consistency is rigorously verified using the time-varying normalized estimation error squared (NEES) defined as $\mathrm{NEES}_k = \mathbf{e}_k^{\mathsf T}\mathbf{P}_k^{-1}\mathbf{e}_k$, where $\mathbf{e}_k$ is the error vector at epoch $k$, and $\mathbf{P}_k$ is the filter's posterior covariance. Finally, empirical tracking accuracy is assessed by computing the absolute error magnitudes of the 8D kinematic and clock states relative to the ground-truth trajectory.

\section{Results and Discussions} \label{sec:results}
\subsection{Fundamental Performance Bounds: Snapshot vs. Recursive}\label{subsec:rec_vs_snap}
\begin{figure}
    \centering
    \includegraphics[width=1\linewidth]{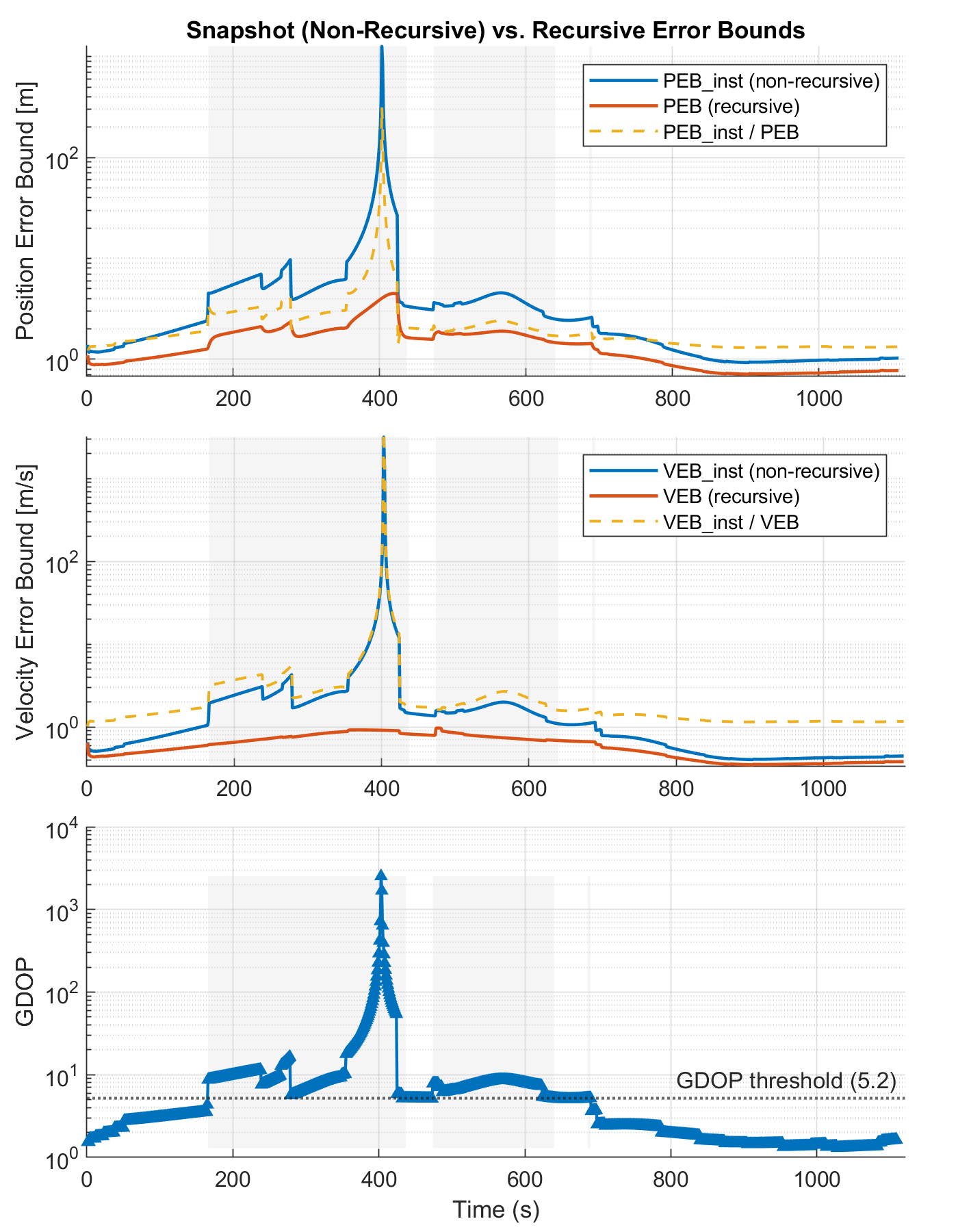}
    \caption{Comparison of the instantaneous (snapshot) and recursive Bayesian performance bounds for position (PEB, top) and velocity (VEB, middle) along the \texttt{Urban03} trajectory, and GDOP (bottom).}
    \label{fig:rec_vs_snap_gdop}
\end{figure}
\begin{figure}
    \centering
    \includegraphics[width=1\linewidth]{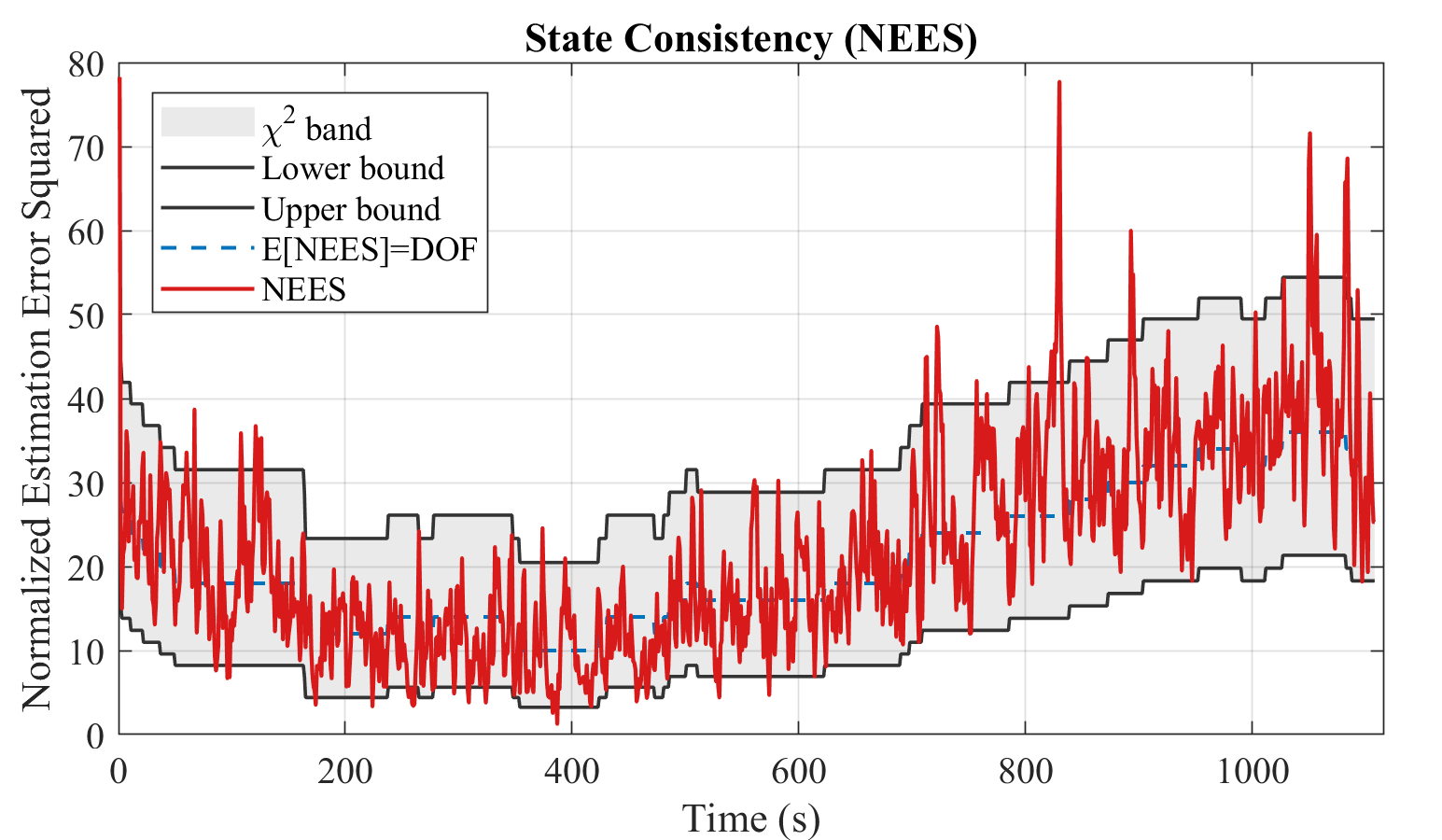}
    \caption{Normalized estimation error squared analysis on the BS filter.}
    \label{fig:bs_nees}
\end{figure}
Fig.~\ref{fig:rec_vs_snap_gdop} compares the snapshot (non-recursive) bounds against the recursive posterior bounds produced by the CRB recursion for both position and velocity and the corresponding geometric dilution of precision (GDOP). It can be seen that the instantaneous (non-recursive) bounds are very sensitive to variation in GDOP, while the recursive bounds are consistently lower and contained under poor GDOP. Two consistent trends are observed. First, when GDOP remains in a moderate range, the snapshot and recursive bounds closely track each other, indicating that the instantaneous measurement information is sufficiently informative such that the prior contributes only marginally. Second, as GDOP increases (poor geometry and/or reduced measurement size or diversity), the ratio between snapshot and recursive bounds grows markedly. In these intervals, the non-recursive bounds inflate rapidly, often by orders of magnitude, because they depend only on the current epoch’s measurement-domain information, which becomes ill-conditioned under weak geometry. In contrast, the recursive bounds remain substantially lower because the Bayesian recursion injects accumulated information from past epochs through the state dynamics and process model. This behavior is most evident near the large GDOP excursion around $t\!\approx\!400$~s, where, as shown in Fig.~\ref{fig:sat_visibility}, the number of visible satellites drops to only four, causing the snapshot bounds to blow-up, while the recursive bounds increase more moderately. This shows that tracking memory becomes even more important when geometry/number of measurements temporarily collapse.

Comparing the position and velocity recursive bounds, it can be seen that the position bound is generally more sensitive to geometry degradation. This can be explained by the usage of the CV process model, which propagates an informative prior on the velocity states directly. In contrast, the position states behave as the time integral of velocity, so any residual uncertainty in velocity accumulates into position. This accumulation can be seen more dramatically during low GDOP, where position states are not properly updated.
\begin{figure}
\begin{subfigure}[]{\linewidth}
    \includegraphics[width=\linewidth]{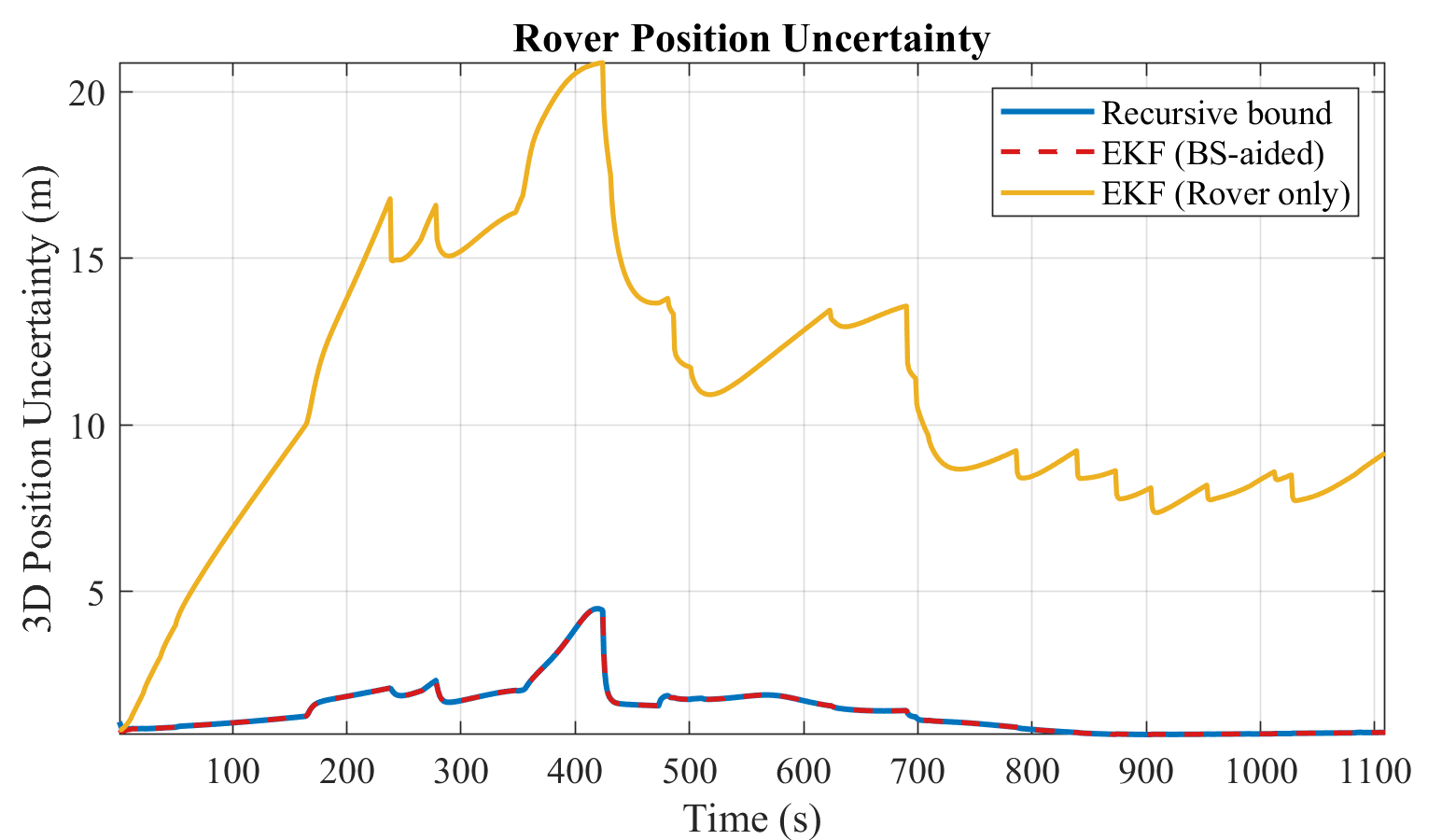}
    \caption{}
    \label{fig:rover_pos_bounds}
\end{subfigure}
\begin{subfigure}[]{\linewidth}
    \includegraphics[width=\linewidth]{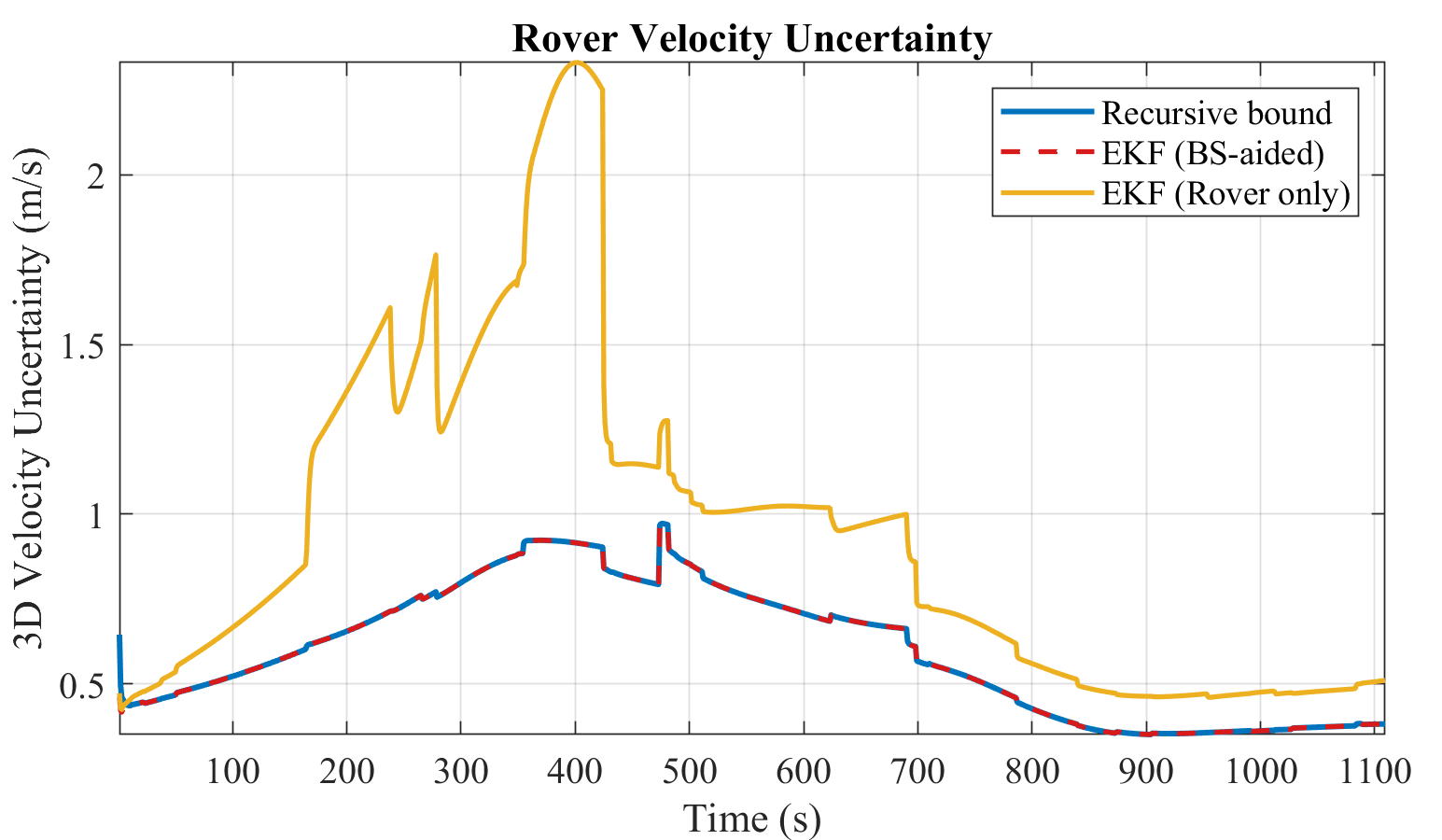}
    \caption{}
    \label{fig:rover_vel_bounds}
\end{subfigure}
\begin{subfigure}[t]{\linewidth}
    \centering
    \includegraphics[width=1\linewidth]{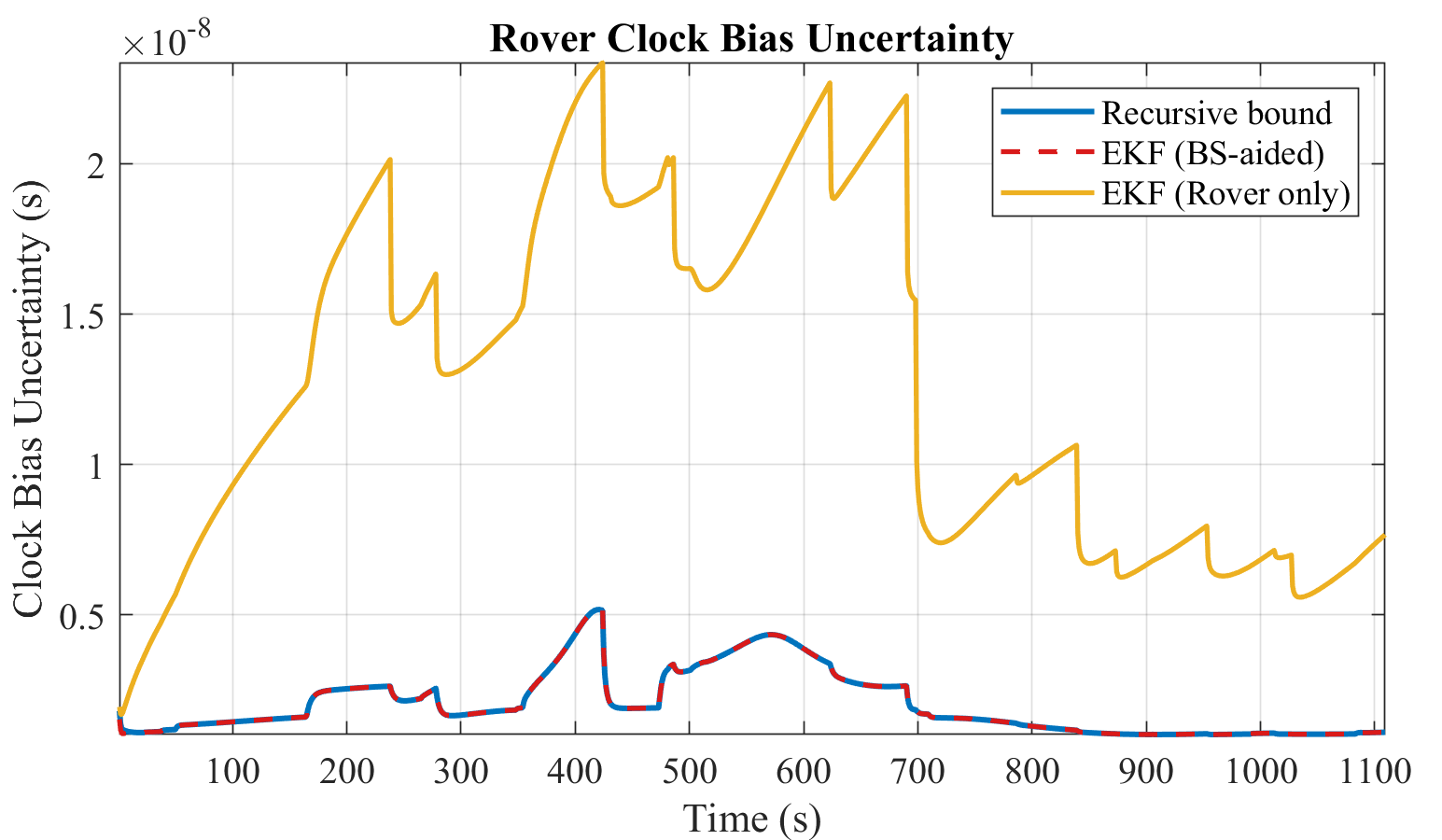}
    \caption{}
    \label{fig:placeholder}
\end{subfigure}
\begin{subfigure}[]{\linewidth}
    \centering
    \includegraphics[width=1\linewidth]{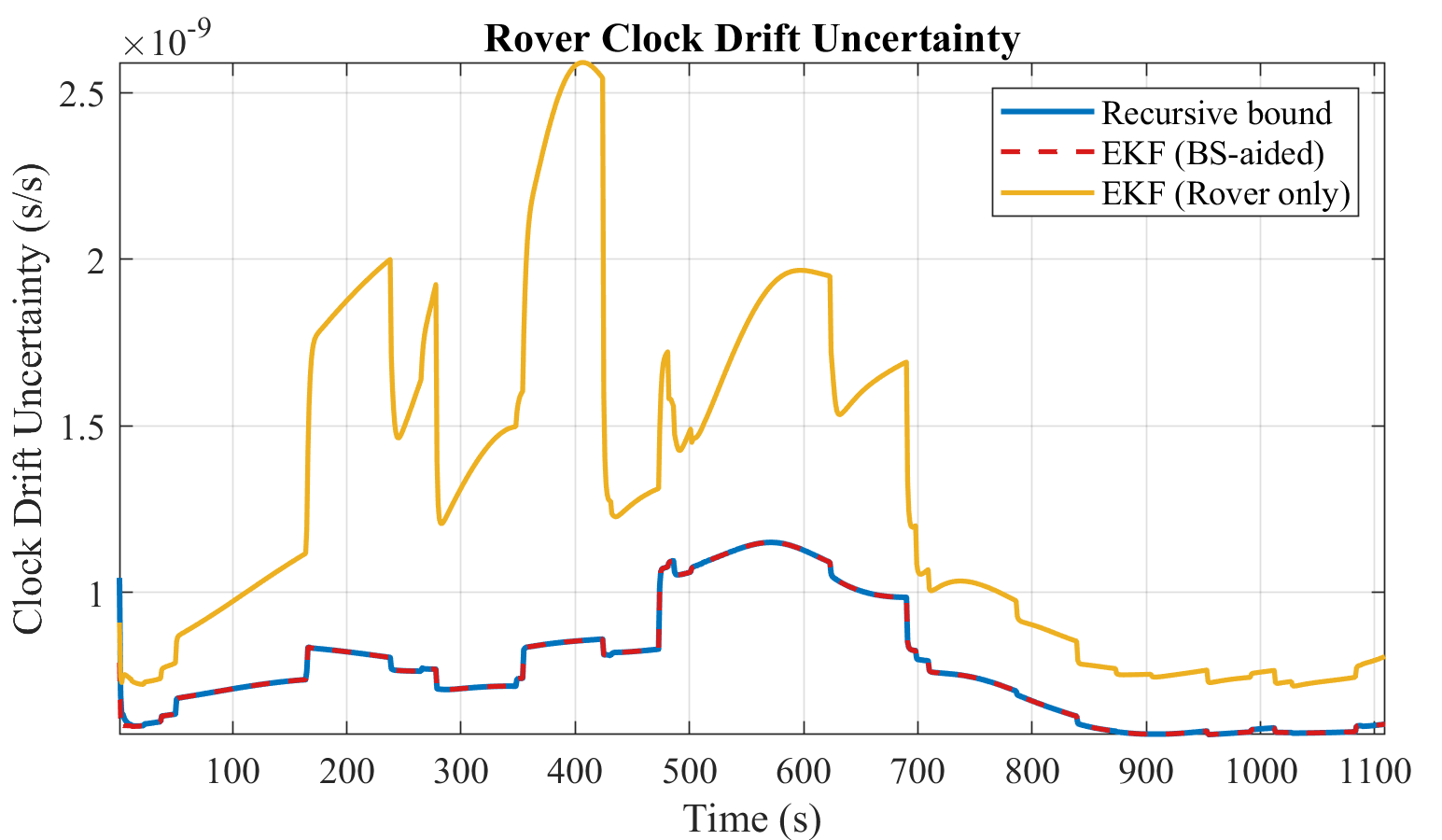}
    \caption{}
    \label{fig:rover_d_bounds}
\end{subfigure}
\caption{Rover EKF posterior uncertainty (1$\sigma$) versus the RBCRB for the (a) 3D position, (b) 3D velocity, (c) receiver clock bias, and (d) receiver clock drift states.}
\label{fig:rover_bounds}
\end{figure}

\begin{figure}[ht!]
    \centering
    \includegraphics[width=1\linewidth]{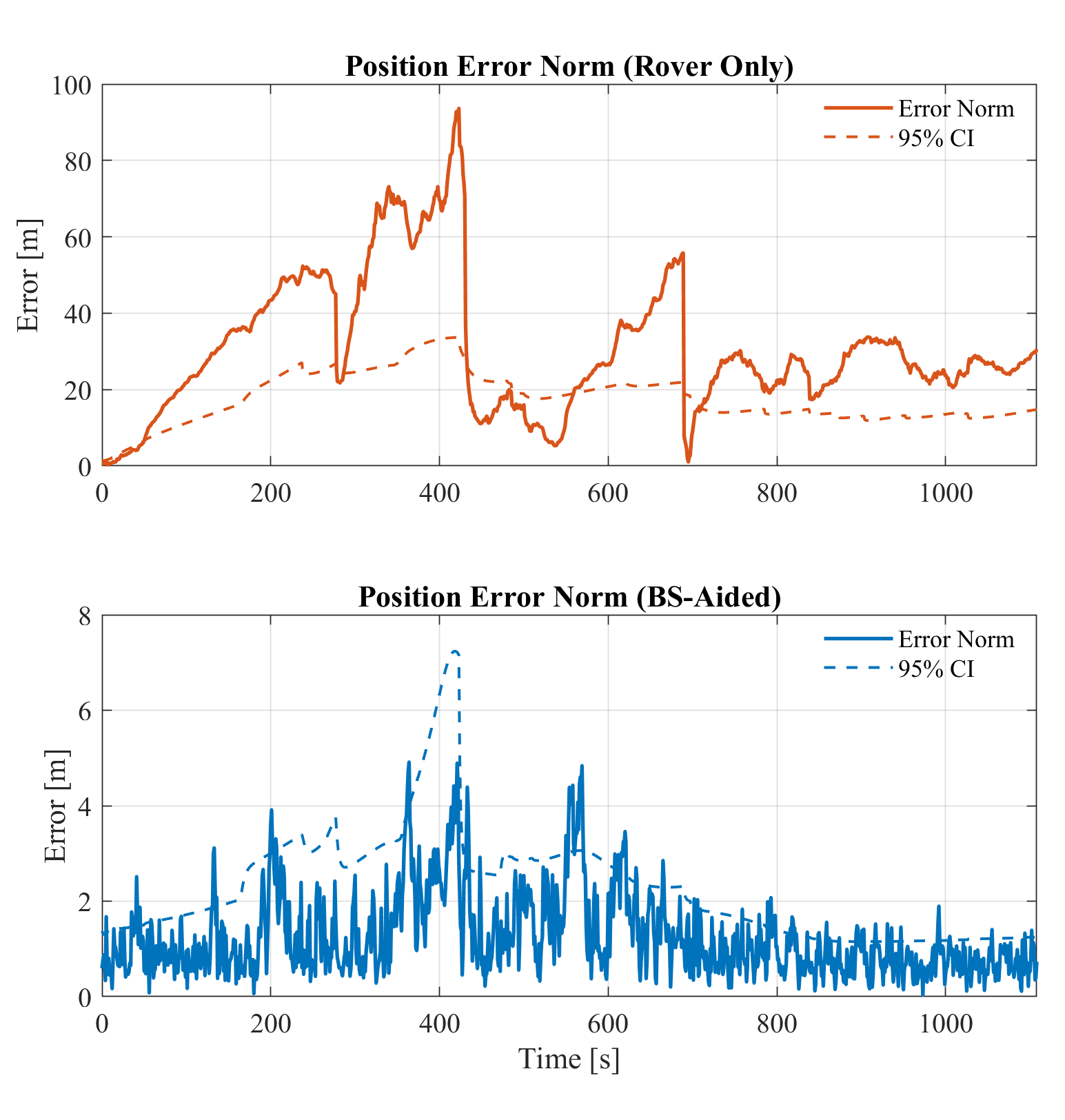}
    \caption{Rover position error magnitude for the rover-only baseline and the BS-aided differential EKF.}
    \label{fig:err_pos}
\end{figure}

\subsection{Statistical Consistency of the Base Station Estimator}
Before assessing rover navigation performance with and without BS assistance, it is essential to verify that the BS clock filter provides a \emph{statistically reliable} covariance. This is because the proposed BS-aided operation injects the BS clock uncertainty into the rover measurement covariance, and any miscalibration at the BS (overconfidence or excessive conservatism) directly biases the effective tuning of measurement covariance at the rover. To this end, we evaluate the BS estimator consistency using NEES analysis, and compare it against the corresponding $\chi^2$ acceptance region at the $95\%$ confidence level, whose bounds vary with the \emph{time-varying} state dimension (i.e. degrees of freedom) (DoF) induced by satellite entry/exit. Fig.~\ref{fig:bs_nees} shows that the BS NEES largely tracks the expected value $\mathbb{E}[\mathrm{NEES}_k]=\mathrm{DoF}_k = 2m_k$ and remains within the confidence band for most time instances, indicating that the reported covariance is broadly representative of the true estimation uncertainty. This behavior is strongly influenced by how newly appearing satellites are initialized in the BS clock filter. In particular, an overly optimistic or otherwise mismatched prior covariance for the entering satellite states can produce abrupt covariance inconsistency at the entry epoch, which is then amplified when satellite entry/exit occurs frequently. Under such conditions, repeated re-initialization events can inject large transients into the filter, inflate the NEES dramatically, and in severe cases drive the estimator toward numerical instability or divergence.

\subsection{Rover Estimator performance with and without BS assistance} \label{subsec:results_rover_vs_bounds}

Having established the BS filter reliability via NEES analysis, we now evaluate the statistical consistency and efficiency of the proposed estimation frameworks. Figs.~\ref{fig:rover_pos_bounds}--\ref{fig:rover_d_bounds} compare the theoretical recursive bounds against the square root of the posterior error variances (standard deviations) extracted directly from the rover EKF covariance matrix $\mathbf{P}_{u,k}$ across the 8D state vector: 3D position, 3D velocity, receiver clock bias, and receiver clock drift.

In the rover-only mode (orange), satellite clock effects remain unmodeled in the rover state and therefore appear as structured measurement uncertainty; this drives substantially larger posterior uncertainties, most notably in position and receiver clock bias. In contrast, the BS-aided mode (red) applies measurement-domain satellite clock corrections and propagates their uncertainty into the rover measurement covariance, leading to a marked reduction in the rover posterior uncertainty and a noticeably closer tracking of the RBCRB envelope.

Additionally, as observed earlier, the temporal structure of the curves reflects the time-varying visibility and geometry of the satellites. During periods of degraded geometry or abrupt satellite set changes, rover-only uncertainty grows rapidly as unmodeled satellite clock terms accumulate; the BS-aided mode remains considerably more stable. 

Crucially, the posterior standard deviations of the BS-aided EKF perfectly overlay the theoretical RBCRB (solid blue line) for all 8 states throughout the entire trajectory, validating the proposed two-stage estimator architecture.
\begin{figure}[t!]
    \centering
    \includegraphics[width=1\linewidth]{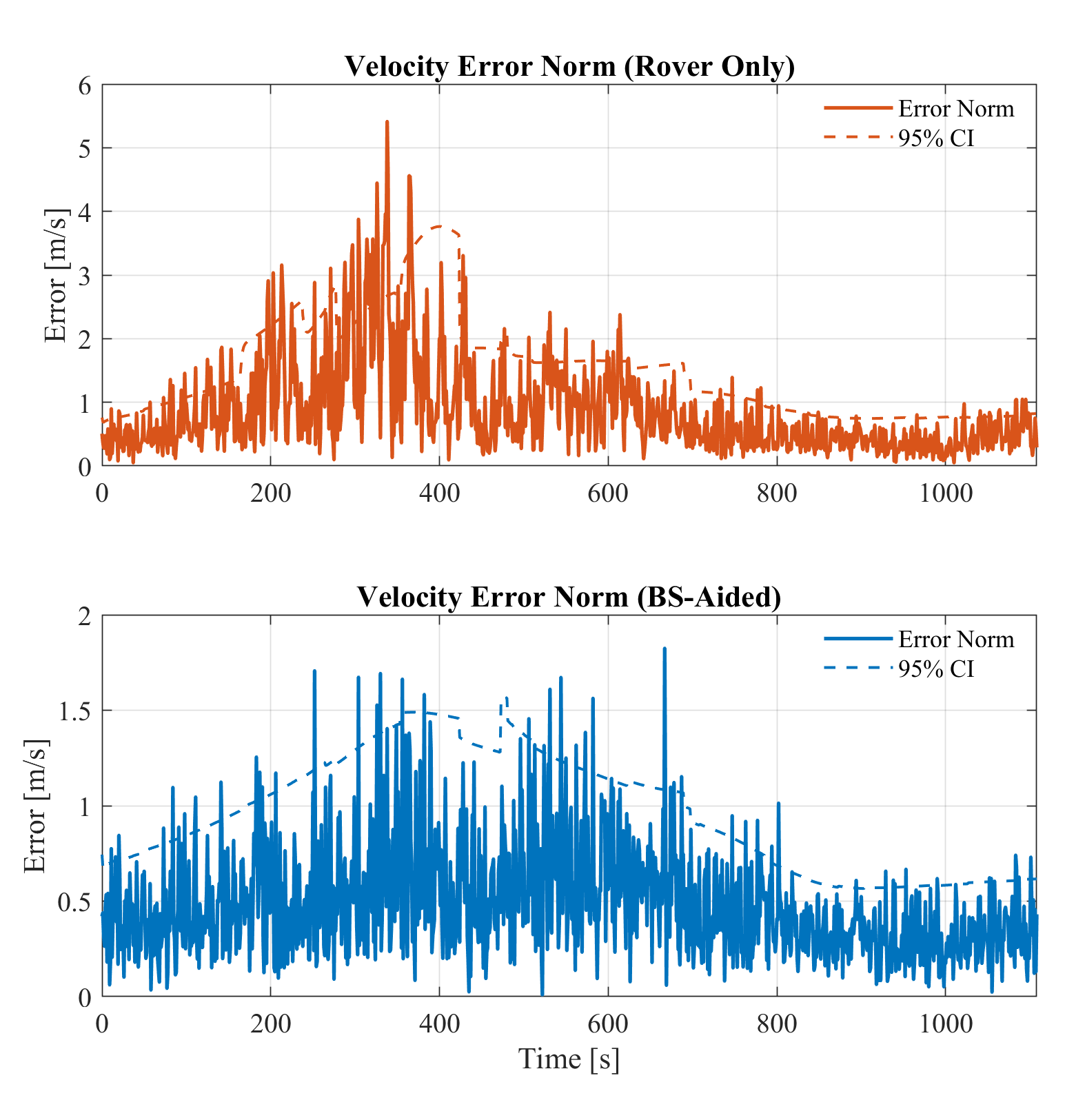}
    \caption{Rover velocity error magnitude for the rover-only baseline and the BS-aided differential EKF.}
    \label{fig:placeholder}
\end{figure}

\begin{figure}[]
    \centering
    \includegraphics[width=1\linewidth]{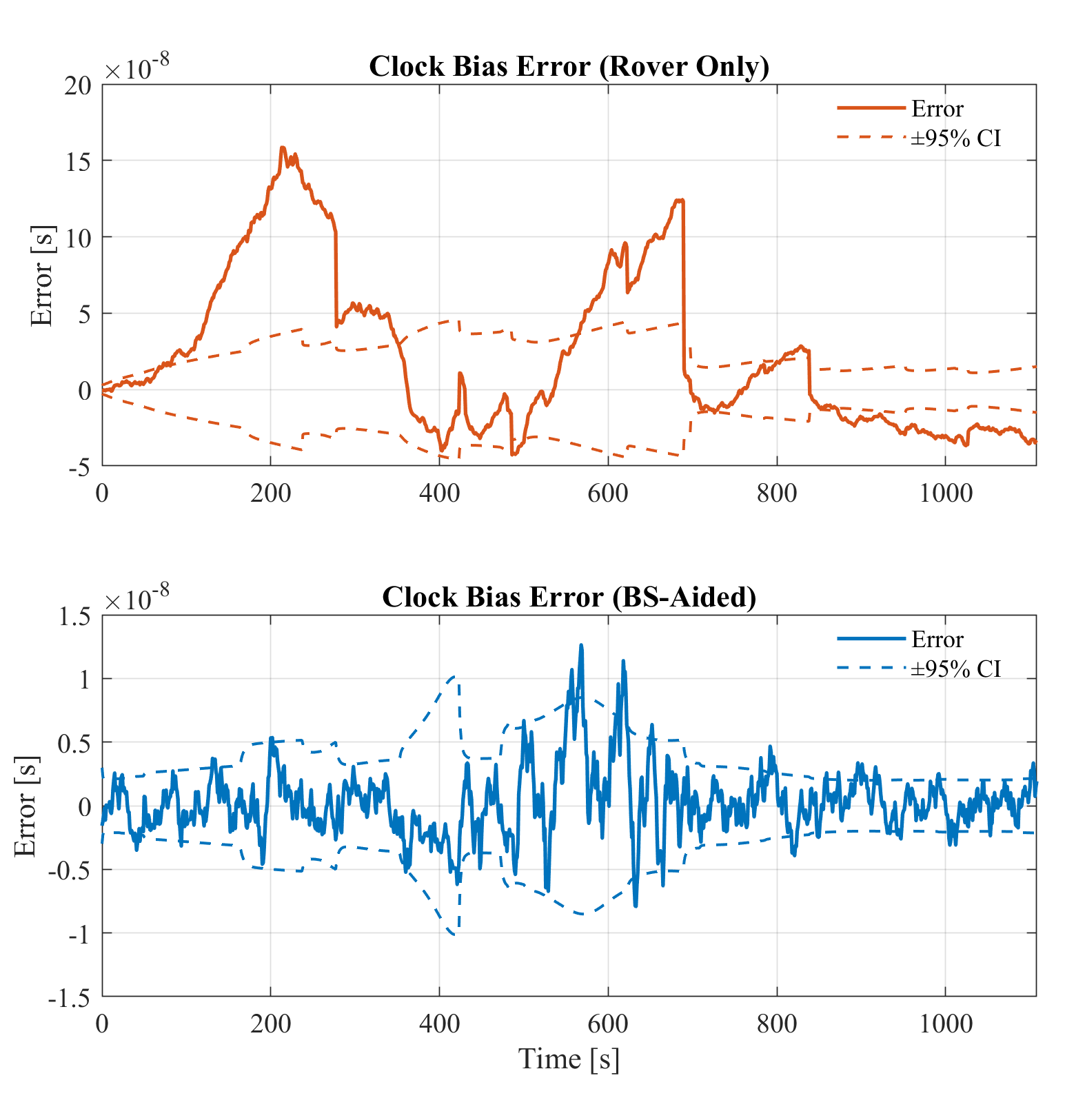}
    \caption{Rover clock bias error magnitude for the rover-only baseline and the BS-aided differential EKF.}
    \label{fig:placeholder}
\end{figure}
\begin{figure}[t!]
    \centering
    \includegraphics[width=1\linewidth]{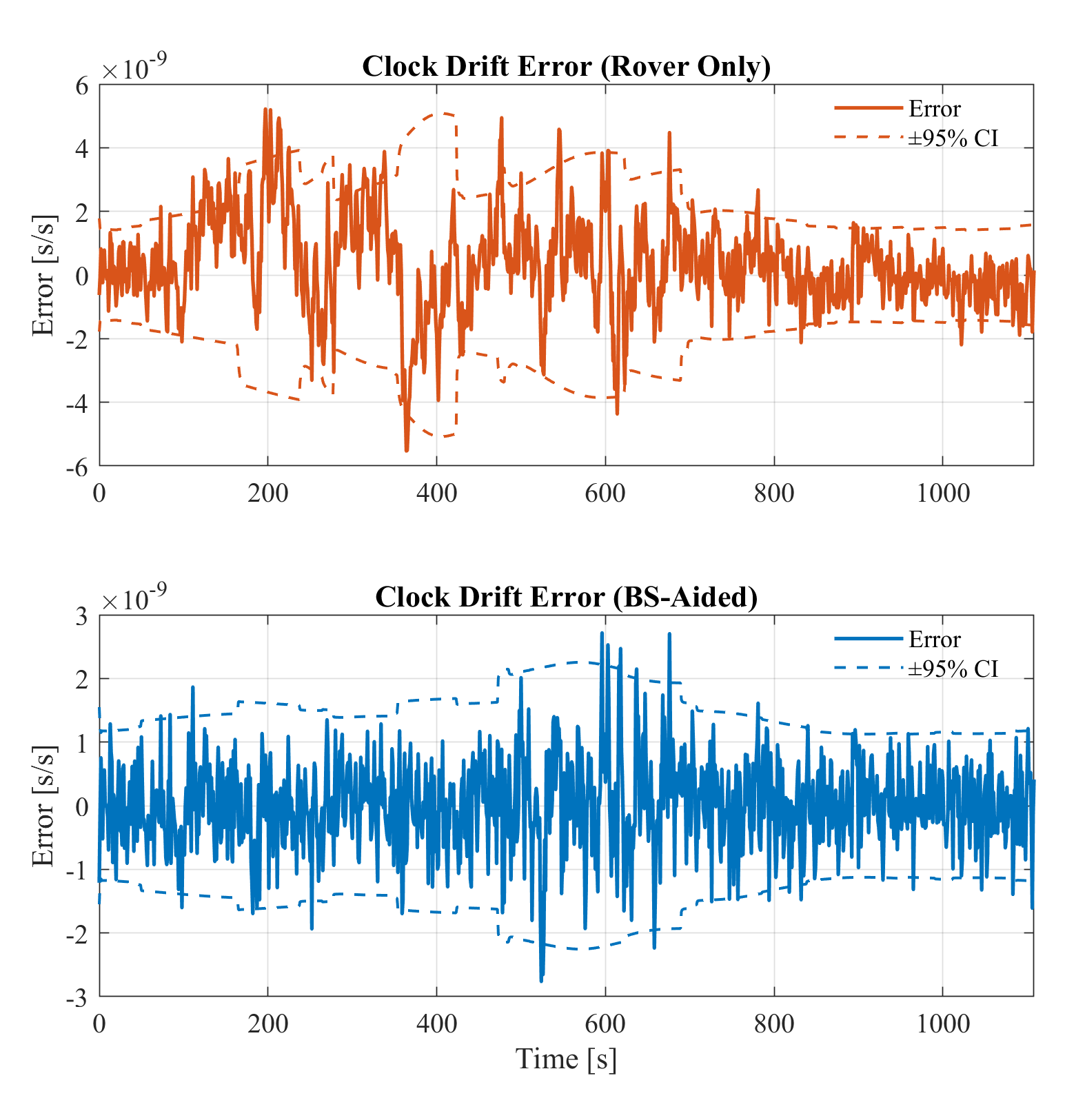}
    \caption{Rover clock drift error magnitude for the rover-only baseline and the BS-aided differential EKF.}
    \label{fig:err_d}
\end{figure}
\subsection{Empirical Tracking Accuracy and Filter Consistency}
Figs.~\ref{fig:err_pos}--\ref{fig:err_d} report the rover estimation errors for position, velocity, clock bias, and clock drift, together with the corresponding pointwise $\pm95\%$ confidence intervals derived from the EKF posterior covariance. In the rover-only baseline, the position and clock-bias errors exhibit pronounced low-frequency excursions and repeatedly violate the confidence intervals, indicating that the filter is effectively absorbing residual satellite timing errors into the accumulating states $(\mathbf{p}_k,b_{u,k})$ and consequently becomes overconfident. In contrast, under BS-aided operation, the clock-bias error is reduced by roughly an order of magnitude and remains largely contained within the confidence envelope, while the position error norm is similarly suppressed and better aligned with its predicted uncertainty. For the rate states, the difference between rover-only and BS-aided is less pronounced: the velocity and clock-drift errors are generally more contained in both modes and show fewer sustained violations, which is consistent with the constant-velocity/clock process model directly regularizing these states and preventing unmodeled timing effects from integrating into large offsets. Overall, these results confirm that BS assistance primarily improves the accuracy and statistical consistency of the rover solution in the states most sensitive to accumulating timing errors, namely position and receiver clock bias.

Finally, large variations can be seen in both the estimation errors and the confidence intervals due to changes in the active satellite set. As satellites enter or exit the visibility set, the measurement information content changes abruptly. Consequently, the uncertainty carried by newly initialized links can temporarily increase innovation magnitudes and produce short-lived error spikes. This highlights the importance of careful satellite association/state registration and principled initialization of newly visible satellites, since high entry/exit rates can otherwise degrade stability and, in extreme cases, lead to divergence.

\section{Conclusion}\label{sec:conclusion}
In this work, a moving rover performs LEO-based positioning and clock tracking in an urban environment using downlink delay and Doppler signals of opportunity. The key challenge is that communication-class LEO satellites do not provide GNSS-like time reference, and their clock bias and drift enter the rover measurements as time-varying nuisance terms; and hence, when these terms are naively appended to the rover state, the resulting model becomes weakly observable under a time-varying visibility set. To address this, we proposed a BS-aided differential architecture operating in two coupled stages. In the first stage, a fixed base station observes the same satellites and tracks per-satellite clock bias and drift using geometry-free delay/Doppler residuals, while explicitly maintaining covariance through satellite entry/exit and index registration. In the second stage, the rover EKF uses measurement-domain corrected delay/Doppler, where the BS clock estimates are applied as per-link corrections, and their uncertainty is propagated into the rover measurement covariance, thereby preserving a compact rover state and enabling consistent fusion.
 
For assessment and benchmarking, we derived the recursive Bayesian Cram\'er--Rao lower bounds that account for the rover motion model and the time evolution of information. Simulation results using a real urban vehicle trajectory and dynamically changing LEO visibility illustrate that BS assistance substantially reduces the dominant error modes in the rover-only baseline.

Future work will extend the proposed differential architecture to consider multipath and NLoS interference prevalent in dense urban environments. Additionally, we plan to investigate multi-BS cooperation, leveraging networked spatial diversity to enhance the geometry of differential corrections and ensure robust, continuous satellite clock tracking across wider coverage areas.

\bibliographystyle{IEEEtran}
\bibliography{References}
\begin{IEEEbiography}[{\includegraphics[width=1in,height=1.25in,clip]{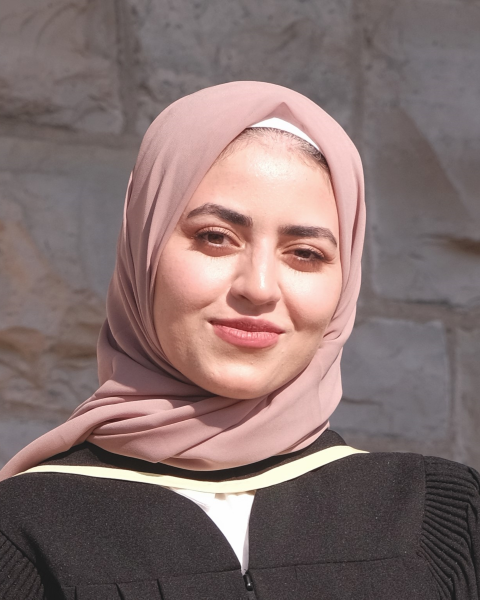}}]{Qamar Bader} (Member, IEEE) received the B.Sc. degree in electrical engineering from Qatar University, Doha, Qatar, in 2016, and the M.A.Sc. and Ph.D. degrees in electrical and computer engineering from Queen's University, Kingston, ON, Canada, in 2023 and 2026, respectively. She is a researcher with the Navigation and Instrumentation Research Lab at the Royal Military College of Canada, RMCC. Her research interests include terrestrial and non-terrestrial positioning and navigation, sensor fusion, and state estimation.
\end{IEEEbiography}
\begin{IEEEbiography}[{\includegraphics[trim=0 700 0 300, width=1in,height=1.25in,clip]{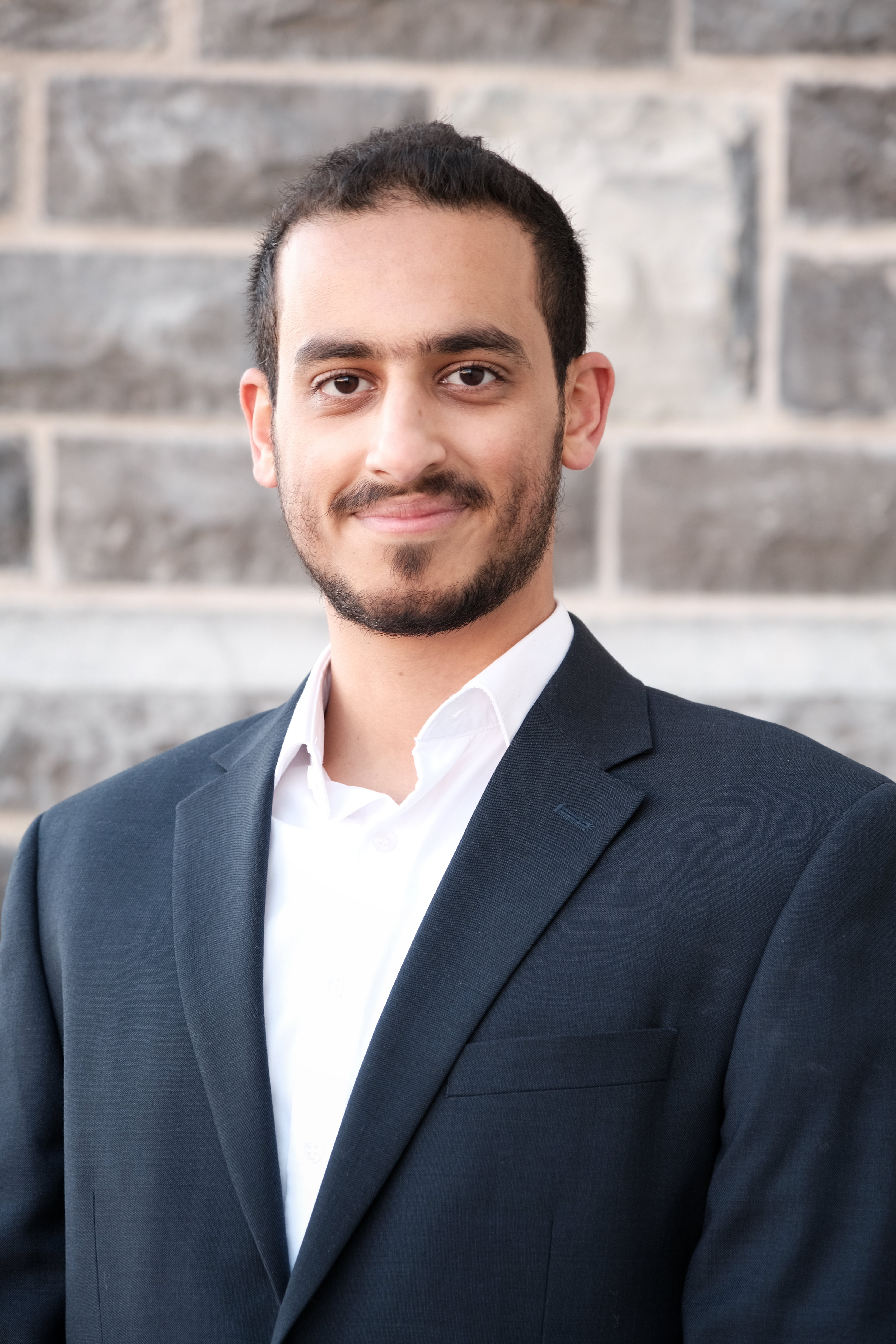}}]{Sharief Saleh} (Member, IEEE) received the B.Sc. and M.Sc. degrees in electrical engineering from Qatar University, Doha, Qatar, in 2016 and 2018, respectively, and the Ph.D. degree in electrical and computer engineering at Queen’s University, ON, Canada, in 2023. He is currently a Marie Skłodowska-Curie Actions (MSCA) Postdoctoral Fellow in the Department of Electrical Engineering at Chalmers University of Technology, Gothenburg, Sweden. His current research interests include 5G/6G positioning in terrestrial and non-terrestrial networks, channel modeling and estimation, and sensor fusion.
\end{IEEEbiography}
\begin{IEEEbiography}[{\includegraphics[trim=0 0 0 0, width=1in,height=1.25in,clip]{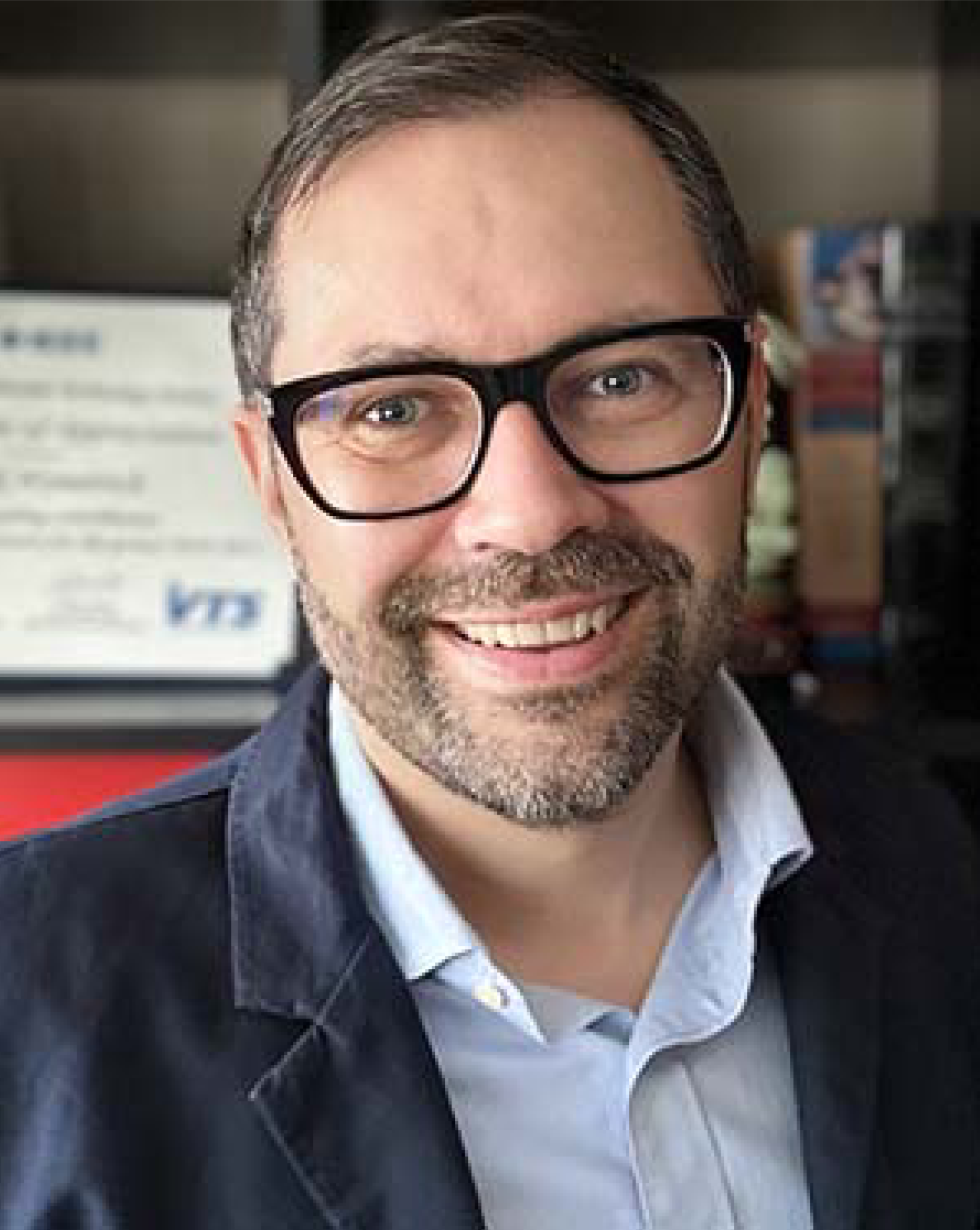}}]{Henk Wymeersch} (Fellow, IEEE) obtained the Ph.D. degree in Electrical Engineering/Applied Sciences in 2005 from Ghent University, Belgium. He is currently a Professor of Communication Systems with the Department of Electrical Engineering at Chalmers University of Technology, Sweden and a Distinguished Visiting Professor at Tsinghua University. Prior to joining Chalmers, he was a postdoctoral researcher from 2005 until 2009 with the Laboratory for Information and Decision Systems at the Massachusetts Institute of Technology. Prof. Wymeersch served as Associate Editor for IEEE Communication Letters, IEEE Transactions on Wireless Communications, and IEEE Transactions on Communications and is currently Senior Member of the IEEE Signal Processing Magazine Editorial Board.  During 2019-2021, he was an IEEE Distinguished Lecturer with the Vehicular Technology Society.  His current research interests include the convergence of communication and sensing, in a 5G and Beyond 5G context. 
\end{IEEEbiography}
\begin{IEEEbiography}[{\includegraphics[trim=0 0 0 0, width=1in,height=1.25in,clip]{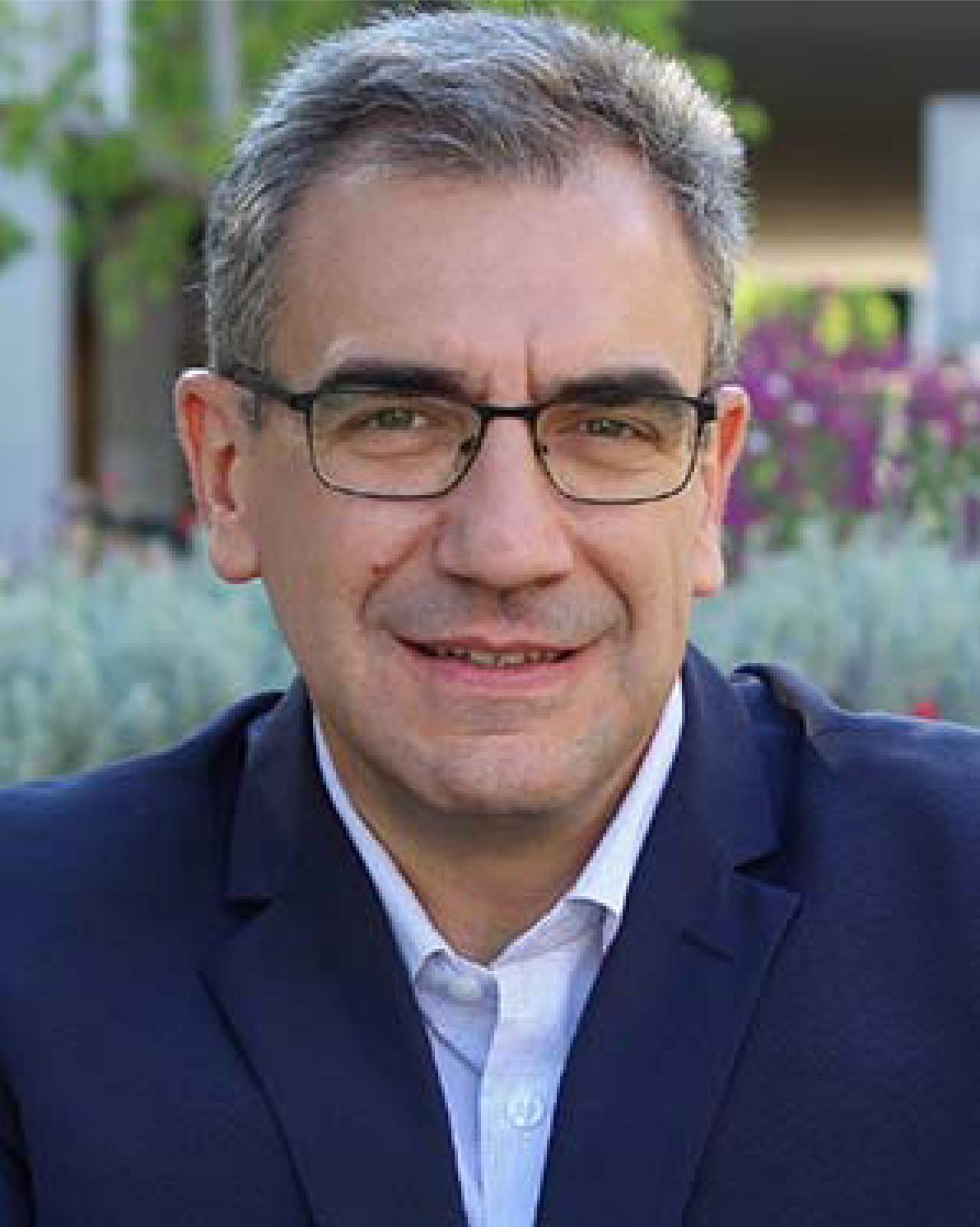}}]{Gonzalo Seco-Granados} (Fellow, IEEE) received
the Ph.D. degree in telecommunications engineering from the Universitat Politècnica de Catalunya, Barcelona, Spain, in 2000, and the M.B.A. degree from IESE Business School, Barcelona, in 2002. From 2002 to 2005, he was with the European Space Agency, where he contributed to the design of the Galileo system. He is currently a Professor with
the Department of Telecommunications, Universitat Autònoma de Barcelona (UAB), Bellaterra, Spain, and also the Director with the Centre for Space Studies and Research. He is also with the Institute of Space Studies of Catalonia and is an ICREA Academia Fellow. In 2015, 2019 and 2022, he was a Fulbright Visiting Scholar with the University of California, Irvine, CA, USA. His research interests include signal processing for GNSS, LEO-PNT, and 5G/6G integrated communications, localization, and sensing. From 2018 to 2025, he was the President of the Spanish Chapter of the IEEE Aerospace and Electronic Systems Society. He is a member of the EURASIP Signal Processing for Multisensor Systems Technical Committee since 2022. He was the recipient of the 2021 IEEE Signal Processing Society’s Best Paper Award.
\end{IEEEbiography}
\begin{IEEEbiography}[{\includegraphics[trim=0 0 0 0, width=1in,height=1.25in,clip]{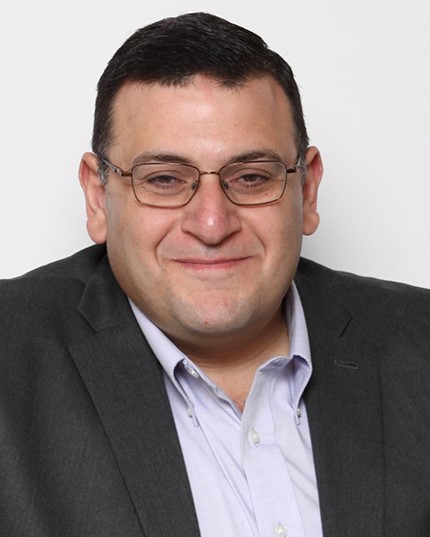}}]{Aboelmagd Noureldin} (Fellow, IEEE) is a Professor and Tier-I Canada Research Chair (CRC) in resilient high-precision positioning and navigation at the Department of Electrical and Computer Engineering, Royal Military College of Canada (RMC), with Cross-Appointment at both the School of Computing and the Department of Electrical and Computer Engineering, Queen’s University. He is also the founding director of the Navigation and Instrumentation (NavINST) research group at RMC, a unique world-class research facility in GNSS, wireless positioning, inertial navigation, remote sensing, and multisensory fusion for navigation and guidance. He holds a Ph.D. in Electrical and Computer Engineering (2002) from The University of Calgary, Alberta, Canada. In addition, he has a B.Sc. in Electrical Engineering (1993) and an M.Sc. degree in Engineering Physics (1997), both from Cairo University, Egypt. He is a professional member of the Institute of Navigation (ION). He published two books, four book chapters, and numerous papers in academic journals, conferences, and workshop proceedings, for which he received several awards. His research led to 13 patents and several technologies licensed to industry in positioning and navigation systems.
\end{IEEEbiography}
\end{document}